\documentclass[a4paper,10pt]{scrartcl}

\usepackage[T1]{fontenc}
\usepackage{lmodern}

\usepackage[pdftex,usenames,dvipsnames]{color}

\usepackage{graphicx}
\usepackage[german,american]{babel}

\usepackage{algorithm}
\usepackage{algorithmic}

\usepackage{subfigure}
\usepackage{wrapfig}
\usepackage{setspace}

\usepackage{amsmath}
\usepackage{amssymb}
\usepackage{amsfonts}
\usepackage{latexsym}
\usepackage{mathtools}
\usepackage{bm}

\usepackage{hyperref}

\newcommand{\IP}{\mathbb{P}}

\newcommand{\cL}{{\mathcal L}}

\newcommand{\cN}{{\mathcal N}}

\newcommand{\cU}{{\mathcal U}}
\newcommand{\cO}{{\mathcal O}}

\usepackage{listings}
\newcommand{\code}[1]{\texttt{#1}}
\lstdefinestyle{python}{
  frame=tb,
  xleftmargin=\parindent,
  language=Python,
  basicstyle=\ttfamily,
  keywordstyle=\bfseries\color{NavyBlue},
  commentstyle=\color{ForestGreen},
  stringstyle=\color{Brown},
  showstringspaces=false,
}
\lstdefinestyle{pythonsmall}{
  frame=tb,
  xleftmargin=\parindent,
  language=Python,
  basicstyle=\small\ttfamily,
  keywordstyle=\bfseries\color{NavyBlue},
  commentstyle=\color{ForestGreen},
  stringstyle=\color{Brown},
  showstringspaces=false,
}
\usepackage{xr}
\newcommand{\Pobs}{\cO}
\newcommand{\bth}{\theta}

\newcommand{\figpath}{figures}

\usepackage[bottom = 4.00cm]{geometry}

\begin{document}

\title{\LARGE\vspace*{-50pt}SPUX Framework: a Scalable Package for Bayesian Uncertainty Quantification and Propagation}
\author{Jonas {\v S}ukys and Marco Bacci}
\maketitle

\begin{abstract}
We present SPUX - a modular framework for Bayesian inference
enabling uncertainty quantification and propagation
in linear and nonlinear, deterministic and stochastic models,
and supporting Bayesian model selection.
SPUX can be coupled to any serial or parallel application written in any programming language,
(e.g.~including Python, R, Julia, C/C++, Fortran, Java, or a binary executable),
scales effortlessly from serial runs on a personal computer
to parallel high performance computing clusters,
and aims to provide a platform particularly suited to support and foster reproducibility in computational science. 
We illustrate SPUX capabilities for a simple yet representative random walk model,
describe how to couple different types of user applications,
and showcase several readily available examples from environmental sciences.
In addition to available state-of-the-art numerical inference algorithms
including EMCEE, PMCMC (PF) and SABC,
the open source nature of the SPUX framework and
the explicit description of the hierarchical parallel SPUX executors
should also greatly simplify the implementation and usage of other
inference and optimization techniques.
\end{abstract}

\section{Introduction}
\label{s:introduction}

For centuries, human intuition and curiosity toward natural phenomena have been
the main driving forces behind the discovery of the fundamental laws of physics,
and behind the formulations of mathematical models capable of describing past and
forecasting future behavior of various complex systems.
In evironmental sciences, in particular, there
are strong justifications, and thus a strong trend, for using stochastic models,
such as stochastic differential equations (SDEs) and individual based models (IBMs),
to simulate the dynamical systems of interest.
Indeed, these models are especially useful when intrinsic uncertainties are present
as in, for instance, the modeling of inter-connected systems such as climate, weather,
ocean and lake dynamics, subsurface ground water flows, hydrological catchments, urban
floods, and ecological communities, to name a few.

In recent years,
as reviewed in \cite{acm-computational-sustainability},
two additional important influencing factors have arisen and have become available to scientists:
considerable computational power and a massive increase in data availability.
The steady increase in computational power has allowed models to reduce their level of approximation
to reality by increasing complexity and/or by achieving faster convergence towards the exact solution.
Concurrently, the recent technological advances in sensing and imaging have initiated
the so-called era of big data, allowing one to complement mechanistic modeling based on first principles
(e.g. conservation laws) and human ingenuity with observational data \cite{kutz2016dynamic}. 

However, the opportunity to use high performance computing (HPC) infrastructures
to enable an efficient coupling of complex models and/or of large data-sets, is posing significant challenges to the
so-called scientific programming and computing practices. Indeed, nowadays users are often
required to run their forward simulations on HPC clusters, while developers need to be able to exploit
different types of parallelism to ensure the feasibility of model calibration and uncertainty propagation.
It is also worth noting, that while for some complex models already a single forward simulation can be computationally expensive,
the statistical inference methodologies for assimilating datasets can be extremely demanding already for models of intermediate complexity.

By building on these necessities, the focus of our contribution is on describing a new software
framework, called SPUX, which stands for "Scalable Package for Uncertainty Quantification in X",
that aims to abstract and simplify the access to modern computing infrastructure for reproducible uncertainty quantification and propagation.
The remainder of this section is dedicated to exposing basic information
about Bayesian inference, which is at the core of SPUX (a more detailed exposure is provided in \autoref{s:concepts}),
and to briefly reviewing similar existing computational suites.

To advance the scientific understanding of complex systems,
statistical inference
techniques such as Bayesian inference \cite{Gelman2014}
can be used for
probabilistic quantification (i.e.~including uncertainties)
of model parameters and (past, present and future) model states,
and for comparing several available models using Bayesian model selection.
Bayesian inference conditions the prior distributions of model parameters and (stochastic model) states
(which probabilistically describe any prior information regarding
model parameter and output values)
on the data to get the corresponding posterior distributions.
For instance, it can be based on the so-called likelihood function for a
given model, which formulates the model as a probability distribution of observations
for given model parameter values (to which model's inputs and outputs are associated), and
on the prior distribution of model parameter values.
Posterior probability distribution of model parameters can then
be inferred from combining such prior knowledge about the model and its parameters with
the likelihood function
(see Bayes theorem and \autoref{s:concepts} for more details).

Historically, the successful application of Bayesian inference for stochastic generative models
with realistic datasets has been hindered by the lack of
efficient sampling techniques for posterior model trajectories and for the
computationally expensive evaluation of the likelihood (as a high-dimensional integration).
The development of methodologies to address such challenges has been an active research topic
in recent years.
Relevant methods include
Particle Filter (PF) estimation (with optional trajectory "smoothing" techniques \cite{smoothing})
coupled with Markov Chain
Monte Carlo (MCMC) sampling - also known as Particle Markov Chain Monte Carlo (PMCMC) \cite{Andrieu2010},
Gibbs sampling - including Conditional Ornstein-Uhlenbeck Sampling (COUS) \cite{reichert_2009_hydmodelunc},
the Approximate Bayesian Computation (ABC) methodologies such as Simulated Annealing ABC (SABC) \cite{albert2015simulated},
Hamiltonian Monte Carlo \cite{HMC_Carlo},
and stochastic variational (Bayesian) inference (SVI) methods \cite{SVI}.

In recent years, the number of computational inference frameworks implementing the above-mentioned
methodologies to study not only deterministic but also stochastic models has been growing considerably.
An attempt to provide a thorough review of those suites
would certainly fall short, unless carried out as a dedicated review,
and is therefore beyond the scope of this contribution.
Instead, we briefly mention the existing approaches that we are aware
of to provide a context and to highlight their specificities, referring the reader to the
individual articles.
%
In particular, the first table overviews UQ suites targeted at "static" stochastic models,
where uncertainty is specified by means of hierarchical Bayesian networks, or incorporated into boundary
(and initial) conditions or forcing terms:
\begin{center}
\begin{tabular}{rcccll}
\hline
Name & Language & Type & Type & Methodologies & Reference\\
\hline
BUGS & R/SAS & partial & specialized & Gibbs sampler & \cite{lunn2009bugs}\\
JAGS & Python/R & partial & specialized & Gibbs sampler & \cite{plummer2004jags}\\
MUQ & C++ & partial & framework & optimization and UQ & \cite{parno2014uncertainty}\\
Pest & proprietary & partial & program & optimization and UQ & \cite{doherty2014pest}\\
STAN & Python & partial & framework & MCMC and HMC & \cite{carpenter2017stan}\\
emcee & Python & partial & specialized & EMCEE sampler & \cite{foreman2013emcee}\\
PyMC3 & Python & partial & framework & optimization and UQ & \cite{salvatier2016probabilistic}\\
UQpy & Python & partial & framework & optimization and UQ & \cite{uqpy}\\
SPOTPY & Python & parallel & framework & optimization and UQ & \cite{houska2015spotting}\\
P4U & C/C++ & parallel & framework & optimization and UQ & \cite{hadjidoukas2015pi4u}\\
Dakota & C++ & parallel & framework & optimization and UQ & \cite{adams2009dakota}\\
Korali & Python/C++ & parallel & framework & optimization and UQ & \cite{korali}\\
\hline
\end{tabular}
\end{center}
%
%
The second table overviews UQ suites additionally supporting
more complex "dynamic" stochastic models (e.g. driven by SDEs and/or SPDEs):
\begin{center}
\begin{tabular}{rcccll}
\hline
Name & Language & Type & Type & Methodologies & Reference\\
\hline
PPF & Java & partial & specialized & Particle Filtering (PF) & \cite{demirel2014ppf}\\
timedeppar & R & partial & specialized & COUS & \cite{reichert_2009_hydmodelunc}\\
EasyABC & R & partial & specialized & ABC methods including SABC & \cite{easyabc}\\
pyro & Python & partial & framework & deep learning, UQ, SVI & \cite{pyro}\\
ABCpy & Python & parallel & specialized & ABC-type samplers & \cite{dutta2017abcpy}\\
LibBi & C++ & parallel & specialized & Particle Filtering (PF) & \cite{murray2013bayesian}\\
PDAF & Fortran & parallel & specialized & Kalman/Particle Filtering (K/PF) & \cite{nerger2005pdaf}\\
SPUX & Python & parallel & framework & EMCEE, PF, PMCMC, SABC & [current]\\
\hline
\end{tabular}
\end{center}
In both tables, "partial" indicates only partial parallelization -
either only the UQ algorithms (usually only the outer loops over independent sampling tasks)
are parallelized (manually or natively) or only parallel external user applications are supported
as models, in addition to standard serial models. Full support (at the time of writing of this
manuscript) for hierarchical parallelization of all, possibly nested,
algorithms (sampler, aggregator for multiple datasets, marginal likelihood estimator, model)
is indicated by plain "parallel".

As evinced by the large list of available frameworks, Bayesian inference
is a field that evolves fast, especially when considering stochastic models.
%
Despite the strong commonalities in the naming,
the target applications of most suites fall into several insignificantly overlapping
problem classes, hindering the possibility of a direct comparison.
All of the established uncertainty quantification suites listed above
provide users access to very sophisticated methodologies and are of great value to the scientific community;
however, all of them also have one or several shortcomings.
Possible defficiencies include
no support for "dynamic" stochastic models or parallel models,
limited choice of inference algorithms (for instance, either only Markov- or ABC-type),
and user interfaces based on a rather technical programming language (C/C++/Fortran).
Unfortunately, among all suites reviewed here and actually supporting full parallelism and "dynamic" stochastic models,
none implement both classes of inference algorithms (MCMC-type and ABC-type)
and, at the same time, provide an interface in high-level programming language (e.g.~Python) for both programming use cases:
coupling a user's model and implementing a new inference algorithm.
In addition, since plain embarrassingly parallel Bayesian inference algorithms
are being outcompeted by more efficient and adaptive, but also more
communication intensive methodologies, modern computing inference suites are required to evolve
to incorporate and offer these new functionalities despite their higher algorithmic complexity.
In contrast to the well-established suites for "static" stochastic models,
a niche for UQ suites with a focus on
"dynamic" stochastic models, easy user interface,
support for different classes of inference methodologies, and full parallelization
was so far relatively empty.
In other words, the above overview provides motivation and sets specific goals
for any new uncertainty quantification framework.

With the SPUX framework we aim to offer a framework that is open to a large class of model structures,
and does not pose any limitations to the programming language.
For instance, models can be serial or parallel,
and can be written in any language (e.g.~Python, R, Julia, C/C++, Fortran, Java),
or can be available just as a binary executable.
We choose Python as programming language for SPUX as it is simple, popular, flexible, and yet suited
to exploit modern HPC architectures by means of, for example, the \code{mpi4py} package \cite{Dalcin2011}.
More specifically, our framework mitigates high computational costs by adaptively distributing model
evaluations (for different parameters, data-sets, and stochastic trajectories) over multiple computational
units in a parallel compute environment. It does so according to a multilevel parallel programming approach,
which allows SPUX to overcome the standard paradigm of map-reduce workflows in favor of a more flexible
design tailored towards algorithmic efficiency, as it is suggested in a recent review \cite{levlibfeld2019fast}.
Indeed, the flexible parallelization paradigm used in SPUX is based on a continuous
management of multiple parallel workers, while their internal states are maintained
remotely to significantly improve the efficiency of complex algorithms, such as the
Particle Filtering method.
SPUX can scale effortlessly from laptops to large parallel compute clusters,
where it is particularly suited.
SPUX already natively supports multiple inference approaches, namely,
Affine Invariant Markov chain Monte Carlo Ensemble with or without Particle Filtering
(with memory-efficient "rejection" particle smoothing \cite{jacob2015}),
Simulated Annealing ABC, and also standard Metropolis-Hastings MCMC. To the best of our knowledge,
SPUX might be the first uncertainty quantification framework that gathers all 
these capabilities in a single implementation.
Finally, we want to mention that the open source nature of the SPUX framework
and the explicit description of the hierarchical parallel SPUX executors should
allow to greatly simplify any additional implementation and usage of other
(existing or future) numerical inference and optimization algorithms for deployment on parallel clusters.

The scope of the manuscript is to present the most recent version (1.0) of SPUX --
a prototype of which was already introduced in the earlier publication \cite{Sukys2017a}.
Theory and numerical methods
for Bayesian inference are briefly reviewed in \autoref{s:concepts}.
The purpose, design specification, and available modular components
of SPUX are introduced in \autoref{s:framework},
while \autoref{s:usage} showcases the framework for a simple example
model. A detailed overview of the most common use case -- coupling
of a scientific application to the SPUX framework, including a
novel proposed adaptive sampling strategy,
is provided in \autoref{s:customization}.
Finally, \autoref{s:parallelization}
describes the design of the SPUX framework based
on the parallel SPUX executors, and the outlook to future
developments is provided in \autoref{s:summary}.

\section{Mathematical concepts and numerical algorithms}
\label{s:concepts}

In this section, we start with a brief review of the mathematical concepts
for the underlying scientific problem addressed by our framework.
In particular, we introduce generic and hidden Markov models and Bayesian
inference for them, followed by several widely used numerical techniques,
and a brief summary for subsequent uncertainty propagation (forecasting) of future model predictions.

\subsection{Generic and hidden-Markov (state-space) models}

Within the scope of this uncertainty quantification framework,
we will consider two classes of predictive models:
a wide class of "generic" models and a specialized class of hidden-Markov models.
\begin{figure}[!ht]
\centering
\includegraphics[width=\textwidth]{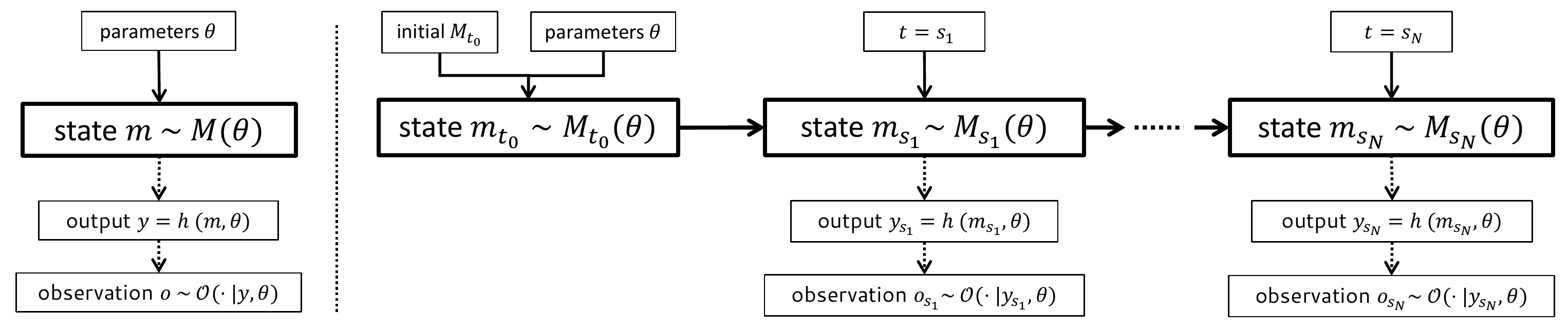}
\caption{Scheme of a generic model (left) and a hidden-Markov model (right),
mapping parameters $\theta$ to the state $m$, output $y$ and observation $o$ of the model $M$.}
\label{f:spux-model-scheme}
\end{figure}

In a generic model $M$, a set of model parameters within a vector $\theta$
is mapped to the model "prediction", as depicted in the left part of \autoref{f:spux-model-scheme}.
We further categorize such model prediction into the full model "state" $m = M(\theta)$,
and its hidden part
(defined by function $h$, for instance, extracting only surface values from a three-dimensional lake model
or accummulating only the number of adult individuals in an ecological community)
as model "output" $y = h (m, \theta)$.
If the model $M$ is stochastic
(e.g. driven by a stochastic process and hence attaining a non-unique state $m$)
then, given a suitable probability measure (denoted by $\IP$),
its state is characterized by a probability distribution $m \sim M(\theta)$,
as a shorthand notation for state $m$ having a probability $\IP (m|\theta,M)$.
The ensemble of all possible "predictions" of such stochastic models
are sometimes also referred to as "trajectories".

Many realistic models have an explicit temporal dimension
(denoted in this manuscript by time $t$)
as depicted in the right part of \autoref{f:spux-model-scheme}.
In such case, for each time $t$, we denote the model by $M_t$,
a specific model state by $m_t$, and the model output as $y_t = h (m_t, \theta)$.
A stochastic time-dependent model $M$ is called a hidden-Markov model,
if, for \emph{any} increasing sequence of times $s_1 < \dots < s_N$,
its corresponding states $m_t \sim M_t (\theta)$ (but not necessarily its outputs $y_t$)
satisfy the Markov property for all $\theta$ and all $1 \leq k \leq N$:
\begin{equation}
\label{eq:markov-property}
\IP(m_{s_k}|m_{s_{k-1}},\dots,m_{s_1},\theta,M) = \IP(m_{s_k}|m_{s_{k-1}},\theta,M).
\end{equation}

In a realistic scenario (to be modeled by model $M$),
the value of the output $y$, which we could refert to as the "exact" or "true" output,
is often not measured completely accurately
during the observation process (independentent of a chosen model type).
In particular, the corresponding data "observations"
$o$ (see \autoref{f:spux-model-scheme})
are instead assumed to follow a probabilistic distribution $\cO (\cdot|y,\theta)$,
sometimes referred to as the observational error model,
potentially also depending on some uncertain parameters,
included within the same vector $\theta$ for simplicity and brevity of the exposition.

For time-dependent models,
the corresponding data observations $o_t$
at "snapshots" $t = s_1, \dots, s_{\bar N}$
are assumed to be mutually independent and each follow a given
$\cO (\cdot|y_{s_n},\theta)$ distribution.
Consequently, the entire observation sequence $(o_{s_1}, \dots, o_{s_N})$
follows a tensorized distribution with a shorthand notation
$\cO (\cdot|y,\theta) = \otimes_{n = 1}^N \cO (\cdot|y_{s_n},\theta)$.
In the following we refer to the observational error model simply by "error".

Finally, hierarchical Bayesian networks
(as examples of "static" probabilistic models introduced in \autoref{s:introduction})
are supported as distributions (see \autoref{s:add:distribution})
for observational error and priors of model parameters and intial model states (see \autoref{s:inference}),
are hence are not incorporated among the "model" concept whithin manuscript.

\subsection{Bayesian inference}
\label{s:inference}

Bayesian inference \cite{Gelman2014} can be used
for statistical quantification (including uncertainties)
of model parameters and (past, present and future) model states
by conditioning the corresponding prior distributions
on the data to get the corresponding posterior distributions.
In particular, for a given model $M$ mapping the parameters vector $\theta$ to
(possibly probabilistic) model state $m \sim M (\theta)$,
the so-called likelihood $L(D|\theta, M) = \IP (D|\theta, M)$
of the model $M$ defines a probability distribution
of observations $D$ for given model parameter values $\theta$.
In addition to the likelihood, initial information about parameters $\theta$
is described probabilistically by the so-called prior distribution $\pi(\theta|M) = \IP (\theta|M)$.
This prior knowledge on the model and its parameters is combined
with the observed data $D$ via the likelihood $L(D|\theta, M)$
to obtain the posterior distribution $\IP (\theta|D,M)$
of model parameters $\theta$:
\begin{equation}
\label{eq:bayes}
\IP(\theta|D, M) =
\frac{L(D|\theta, M) \pi(\theta|M)}{\IP(D|M)}
\propto
L(D|\theta, M) \pi(\theta|M),
\end{equation}
where the Bayesian model evidence term $\IP(D|M)$, useful for model selection, is independent of the parameters $\theta$.

For a deterministic model $M$,
model output $y = h(m, \theta) = h(M (\theta), \theta)$
can be obtained for any arbitrary $\theta$,
and hence the likelihood can be evaluated explicitly by
applying the error,
i.e. $L(D|\theta,M) = \cO (D | y,\theta)$.
In addition to the posterior distribution $\IP(\theta|D, M)$ of model parameters $\theta$
given by \eqref{eq:bayes},
the posterior distribution $\IP(m|D, M)$ of model states $m$
is given directly by propagating $\IP(\theta|D, M)$ through model $M$
using the procedures described in later sections.

For a stochastic model $M$,
a conditional (on $\theta$) prior distribution $\pi(m|\theta,M)$ of the model states $m$
is also required
(for simplicity, the same notation $\pi$ is used
for the prior distributions of model parameters $\theta$ and of model states $m$).
For instance, in a time-dependent stochastic model with state $m_t \sim M_t (\theta)$,
a prior distribution $M_{t_0}(\theta)$
of the initial model state $m_{t_0}$ needs to be specified (possibly conditional on $\theta$),
and then the prior distribution of the later model states $m_t$ at $t > t_0$
is determined by propagating $M_{t_0}(\theta)$ through model $M$.
Given $\pi(m|\theta,M)$, the conditional (on $\theta$) posterior distribution $\IP(m|\theta,D, M)$
of the (stochastic) model state $m$ can be
inferred jointly with the posterior distribution $\IP(\theta|D, M)$ of model parameters $\theta$
by evaluating their joint posterior $\IP(\theta,m|D, M)$:
\begin{equation}
\label{eq:bayes-states}
\IP(\theta,m|D, M) = \frac{\IP(D|\theta,m, M) \pi(\theta,m|M)}{\IP(D|M)}
\propto
\IP(D|\theta,m, M) \pi(\theta|M) \pi(m|\theta,M).
\end{equation}
Note, that marginalization of \eqref{eq:bayes-states} over stochastic model states $m$
recovers \eqref{eq:bayes}, where
the evaluation of the likelihood $L(D|\theta,M)$ entails a marginalization
over all possible model states $m \sim M (\theta)$:
\begin{equation}
\label{eq:L-marginal}
L(D|\theta, M) = \int \IP (D|\theta,m,M) \pi (m|\theta,M) dm.
\end{equation}
In the remaining of this manuscript
the dependence of $L(D|\theta, M)$ on prior model states distribution $\pi (m|\theta,M)$
will be understood implicitly via the dependency on model $M$ which is assumed to provide
a prior distribution $M_{t_0}(\theta)$ for the initial model state $m_{t_0}$.
The information of the prior distribution for later model states $\pi (m_{t>t_0}|\theta,M)$
is usually incorporated as a specific evolution structure within the model $M$ and hence
is also implicitly taken into account as a dependency for likelihood $L(D|\theta, M)$.

\subsection{Numerical methods for Bayesian inference}
\label{s:sampling}

\begin{figure}[!ht]
	\centering
	\includegraphics[width=\textwidth]{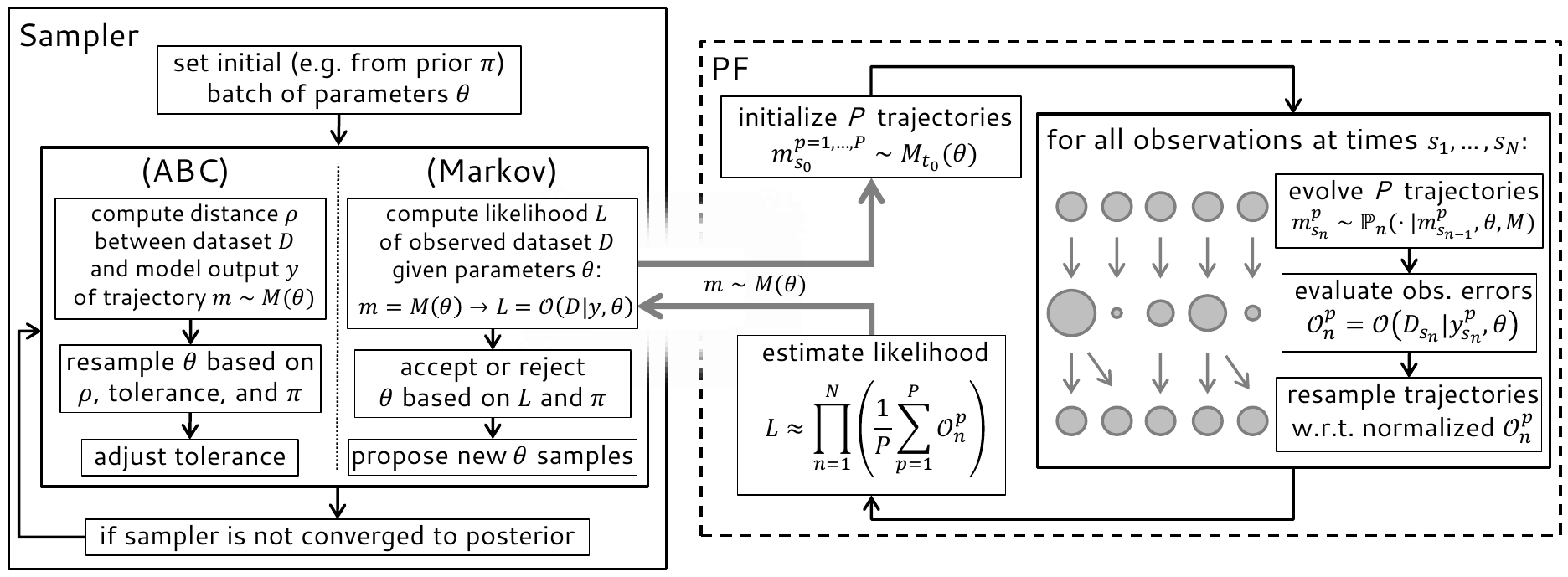}
	\caption{Left: a simplified flowchart of the numerical posterior sampling algorithms
	of the Markov- and ABC-types. Right: a simplified flowchart of the Particle Filter
	(PF) algorithm sampling posterior trajectories (states) of a stachastic model
	by filtering them w.r.t. dataset and the observation error model
	and estimating the likelihood defined as an integral in \autoref{eq:L-marginal}.
	Note, that ABC-type samplers estimate posterior of model parameters (but not states).}
	\label{f:spux-sampler-pf}
\end{figure}

Usually, Bayesian inference cannot be solved analytically for posteriors $\IP(\theta|D, M)$,
and in the case of stochastic model $M$,
usually not even for likelihood $L(D|\theta, M)$ in \autoref{eq:L-marginal}
and hence also not for posterior of model states $\IP(m|D, M)$.
Therefore, numerical methodologies have been developed
to sample from the posterior distribution of model parameters $\theta$
(and states $m$, if the model is stochastic),
obtained by running numerous corresponding simulations of the model $M$.
Existing non-intrusive methods include
the Metropolis or Metropolis-Hastings Markov chain Monte Carlo
(Markov-type for short) \cite{Gamerman2006,Gelman2014,Hastings1970,Metropolis1953},
Gibbs-type samplers for modifying one parameter at a time \cite{reichert_2009_hydmodelunc},
and the Approximate Bayesian Computation (ABC-type for short) \cite{albert2015simulated}.
Alterantive partially intrusive (but usually faster) methods include
Hamiltonian Monte Carlo (HMC-type for short)
and Variational Bayesian Inference (SVI-type for short),
based on exact analytical solution to an approximation of the posterior.
Sampling of posterior model states $m$ in stochastic models can be achieved, for instance,
using
Conditional Ornstein-Uhlenbeck Sampling COUS \cite{reichert_2009_hydmodelunc} within Gibbs-type samplers,
Particle Filtering \cite{Andrieu2010} within Markov-type samplers,
or by recasting model $M$ to consider all its states $m$ as parameters as well \cite{HMC_Carlo} in HMC.
ABC-type methods require minimal model structure restrictions
and are able to reliably sample from model parameters posterior $\IP(\theta,|D,M)$,
but are often very inefficient in sampling posterior of model states $\IP(m|\theta,D,M)$,
since the high-dimensional model output $y$ is usually compressed to a low-dimensional sufficient summary statistic $\mathcal{S}(y)$.
HMC and Variational Inference methodologies usually have the best performance,
but are intrusive (problem reformulations and/or derivatives are required).
Next, we briefly describe Markov-type and ABC-type samplers,
depicted by simplified algorithm flowcharts in \autoref{f:spux-sampler-pf}
and already available within the SPUX framework.

\subsubsection{Markov-type sampling}
\label{s:sampling-markov}

In the Markov-type sampling, samples from the posterior
distribution of model parameters are generated iteratively.
In each iteration, model parameters are proposed,
for which the prior and the likelihood are evaluated
(up to an arbitrary factor) by considering model output (if needed).
These are
then either accepted or rejected (in the latter case, the parameters
from the previoius step are kept). Acceptance or rejection is based
on the ratio of current and previous posterior density estimates (i.e. the
product of likelihood and prior as evinced from Bayes theorem).
In the ABC-type sampling, the initial set of model parameters
is drawn from the prior distribution. Then, an iterative
procedure consist of multiple tolerance steps (converging to zero),
evaluating the distance metric between the model output
and the observations (data), re-drawing
a subset of the model parameters and accepting part of them
based on the distance and the adjusted tolerance,
this way gradually tuning the initial
prior model parameter samples to the posterior model parameter samples.

\subsubsection{Posterior trajectories sampling and likelihood estimation for stochastic models}
\label{s:sampling-markov-stochastic}

For stochastic models, in addition to posterior distribution of the model parameters $\theta$,
sampling of the posterior distribution of model states $m$ is also required.
In particular, the estimation of the (marginalized) likelihood \eqref{eq:L-marginal}
required in the Markov-type samplers is most often estimated numerically.
Nonlinear filtering numerical schemes, such as Particle Filter (PF) with or without smoothing
(also known as Particle Markov Chain Monte Carlo (PMCMC) technique \cite{Andrieu2010})
or a (Seamless) multi-level (Ensemble Transform) Particle Filter ((S)ML(ET)PF) \cite{doi:10.1137/16M1102021}
can be employed.
Any Markov-type sampler can be combined with the (ML)PF
method for likelihood estimations,
where the hidden-Markov structure of the underlying stochastic model $M_t$
is exploited for efficient marginal likelihood approximations
using time-series observations \cite{Andrieu2010,Kattwinkel2017}.
In particular, for observations $D$ consisting of time-series data $D = \{D_{s_n} : n = 1, \dots, N\}$
at time snapshots $t = s_1, \dots, s_N$,
the marginal likelihood in \eqref{eq:L-marginal}
can be rewritten (using the Markov structure, see \cite{Kattwinkel2017}) as
\begin{equation}
\label{eq:L-markov}
L (D | \bth, M) = \int
\pi \left(m_{s_0}|\bth,M\right)
\prod_{n=1}^N \IP_n \left(m_{s_n}|m_{s_{n-1}}, \bth,M\right)
\Pobs (D_{s_n}|y_{s_n}, \bth)
dm_{s_0} \dots dm_{s_n}.
\end{equation}
Here, for given parameters $\bth$, probability distributions
$\IP_n (m_{s_n}|m_{s_{n-1}}, \bth)$
characterize random model state vector $m_{s_n}$
given previous state $m_{s_{n-1}}$, representing
propagation of a given model state $m_{s_{n-1}}$ to the next state $m_{s_{n}}$,
The observational likelihood is evaluated using the (abbreviated) error
$\Pobs_n (D | y, \bth) = \Pobs (D_{s_n}|y_{s_n}, \bth)$
for model output $y_{s_n} = h (m_{s_n},\theta)$
and provides a probabilistic model for the data observation process \cite{Kattwinkel2017}.
The PMCMC algorithm \cite{Andrieu2010,Kattwinkel2017} uses PF to provide
an unbiased statistical estimate of the proposal parameter marginal likelihood
$L (D | \bth, M)$ with structure given in equation \ref{eq:L-markov}.
As depicted in Figure \ref{f:spux-sampler-pf} (right),
the numeric approximation of equation \ref{eq:L-markov}
involves sampling model trajectories ("particles") in terms of model states $m^p$ (with $p = 1, \dots, P$)
of the underlying model $M$ with parameters $\bth$.
At each measurement time $s_n$ in the observations time series,
model simulations are paused and all particles are re-sampled (bootstrapped)
according to their (abbreviated) observational likelihoods
$\Pobs_{n}^p (D|\bth) = \Pobs_n (D | y^p,\bth)$ with $y_{s_n}^p = h (m_{s_n}^p,\theta)$.
Such periodic re-sampling increases algorithmic complexity
due to the required destruction and replication of existing particles,
however, provides an efficient way of sampling
"intermediate" posterior model states
(i.e.~$\IP(m_{s_n}|\theta,D_{s_1,\dots,s_n},M)$ conditioned only
on the partial dataset $D_{s_1,\dots,s_n}$ up to the filtering time $s_n$).
At the end of the PF,
an unbiased estimate $\hat L(D | \bth, M)$ of marginal likelihood $L(D | \bth, M)$
as in equation \ref{eq:L-markov} is evaluated by
\begin{equation}
\label{eq:L-PF}
L (D | \bth, M) \approx
\hat L (D | \bth, M)
=
\prod_{n=1}^N
\left(\frac{1}{P}\sum_{p=1}^P
\Pobs^p_n (D | y, \bth) \right)
=
\prod_{n=1}^N
\left(\frac{1}{P}\sum_{p=1}^P
\Pobs (D_{s_n} | y^p_{s_n}, \bth) \right).
\end{equation}
In implementation, the evaluation fo $\hat L(D | \bth, M)$
is performed in log-scale to mitigate numerical roundoff errors.
The accuracy (namely, the variance) of the PF likelihood estimate clearly depends
on the number of used particles $P$. At the initial burn-in stage,
the sampling acceptace procedure is often dominated by the low likelihood values
and hence the inaccuracy of the PF estimator is of secondary importance.
However, when sampler is converged towards the posterior,
a larger number of particles is prefered to ensure low relative approximation error in likelihood estimator.
In supplementary \autoref{s:pf-adaptivity},
we describe an adaptive procedure to automatically set the number of particles
throughout the sampling procedure based on the feedback
containing historical estimator accuracies and parameters fitness.
To guarantee the convergence of the posterior, the particle adaptivity is
"locked" after the specified period of sampling, which should be smaller
or equal to the burn-in phase.

Additional methodologies, often refered to by "smoothing" \cite{smoothing},
are often employed
to obtain "smoothed" trajectories (conditioned on the entire dataset $D$)
from posterior model states $\IP(m|\theta,D,M)$.
One way to achieve this is to resample already available "intermediate" posterior trajectories
(conditioned only on the partial dataset $D_{s_1,\dots,s_n}$).
In particular, a simple yet very computationally efficient (w.r.t. to both runtime and memory usage)
"rejection" based smoothing (see sections 2.3 and 5 \cite{smoothing} and \cite{jacob2015})
sequentially iterates $s_n$ from $s_1$ to $s_N$ to generate
trajectories from posteriors $\IP(m_{s_1},\dots,m_{s_n}|\theta,D_{s_1,\dots,s_n},M)$
by an additional re-filtering step (after the main PF filter step only for $s_n$)
for all preceeding snapshots $s_1, \dots, s_{n-1}$ as well.
Note, that such "rejection" smoothing, unlike the PF filter for likelihood estimation,
is prone to particle degeneracy
(i.e.~collapses to a single trajectory for each sample of posterior parameter $\theta$)
for $n \ll N$ \cite{smoothing}
and hence should be used with care for non-illustrative purposes.
More sophisticated (but also more computationally expensive)
techniques to prevent such particle degeneracy
have been also reviewed in \cite{smoothing}.

\subsubsection{Approximate Bayesian Computation}
\label{s:sampling-abc}

If the model $M$ does not have a hidden-Markov structure,
or if the error $\cO(\cdot|y,\theta)$ is not explicitly available
as a probabilistic distribution
(i.e.~only a direct sampling of $o = y = h (m,\theta)$ by sampling $m$
from a probabilistic distribution $M (\theta)$ is possible),
then the efficient numerical likelihood estimation methods from
\autoref{s:sampling-markov-stochastic} cannot be directly applied.
Note, that if data $D$ consists only of a single observation,
there are obviously no efficiency gains in using the (temporally) adaptive
likelihood estimation using Particle Filtering.

In such cases, a more general (but potentially less efficient due to the lack of adaptive temporal filtering)
Approximate Bayesian Computation (ABC) methodology \cite{albert2015simulated} can be used to sample from the posterior distribution,
without requiring the evaluation of the likelihood as in \autoref{eq:L-marginal}.
In particular, in Boltzmann-type ABC methods,
the joint posterior of model parameters and states is approximated
by the following family of distributions
\begin{equation}
\label{eq:abc}
\IP_\tau (\theta, m | D, M)
= \frac{1}{Z(\tau)} L (o|\theta, M) \pi (\theta | M) e^{- \frac{\rho (D, o)}{\tau}}
\quad\text{where}\quad
o \sim \cO(\cdot|y,\theta),
\quad
y=h(m,\theta)
\end{equation}
and where $1/Z(\tau)$ is a normalization factor,
$\tau$ is the selected tolerance level,
and $\rho(D, o)$ measures how close the observational dataset $D$
is to the model observation $o$.
If the error $\Pobs$ is not available explicitly,
$o = y = h (m,\theta)$ is used instead.
Given distance $\rho$,
an initial tolerance level $\tau_0$,
and an initial distribution (usually prior $\pi (\theta|M)$),
ABC-type samplers use a sequence of tolerances $\tau_i \to 0$
to generate a sequence of approximations to $\IP (\theta, m | D, M)$.
In the following, dependencies of a particular distance value $\rho (D, o)$
on the parameters $\theta$ and the model $M$ (and, if available, also of the error $\Pobs$)
are explicitly represented by using a supplementary notation $\rho (D|\theta,M)$ for $\rho (D, o)$
with a realization for $o$ obtained from $\theta$ and $M$ as described above.

\subsection{Uncertainty propagation and forecasting}
\label{s:propagation-forecasting}

For time-dependent models $M (\theta) = M_t (\theta)$,
once the joint posterior distribution $\IP(\theta, m_{[t_0, T]}(\theta) | D, M)$
of model parameters and states (up to the last snapshot time $T = s_N$) is available,
the distribution of the "future" (forecast)
model states $\IP(m_{[T,\infty)}(\theta) | D, M)$ is
given by propagating $\IP(\theta, m_T(\theta) | D, M)$
to the "future" times $t > T$ using the model $M$.
In practice, this is achieved with a Monte Carlo (MC) sampler,
where samples from $\IP(m_{[T,T_F]} | \theta, D, M)$
are obtained by sampling from $\IP(m_T | \theta, D, M)$
and propagating them for $t \in [T,T_F]$ for some future time $T_F > T$.

Such an MC sampler can also be used for somewhat less difficult
direct propagation of uncertainty from prior distributions,
when dataset is not available.
For very challenging priors,
Markov-type samplers from \autoref{s:sampling-markov}
can be employed by using prior density instead of the likelihood.

\section{SPUX framework}
\label{s:framework}

In this section we introduce the SPUX framework, 
focusing on the purpose and design specification in \autoref{s:specification},
available modular components and built-in services in \autoref{s:components},
and parallelization capabilities in \autoref{s:parallelization-scheme}.
A collection of continously updated current and past examples
of SPUX applications is illustrated at the end, in \autoref{s:gallery}.

\subsection{SPUX purpose and design specification}
\label{s:specification}

The purpose of the SPUX framework is to provide a seamless high-level interface
to perform Bayesian inference with a free choice of methodologies, algorithms,
and computational environments.
To achieve such flexible customization and effortless adaptivity,
the SPUX framework harnesses the powerful dynamic typing and runtime polymorphism
offered by the modern Python programming language,
which has recently become one of the most popular programming languages for scientific computing.
In essence, the SPUX framework is a collection of carefully
selected, designed and optimized modular components.
The modularity of SPUX components extends beyond the conventional restrictive
patterns and instead follows a "duck typing" design philosophy \cite{duck}, namely,
the suitability of an object to perform a function is not determined by
the object's type, rather by the support of certain methods and properties
by the object itself. An overview of the basic key components currently implemented
in SPUX, each with the purpose of representing a particular mathematical
concept as introduced in \autoref{s:concepts}, together with
available specific numerical methods for each component type, is provided
in the table below:
\begin{center}
\begin{tabular}{c|r|c|l}
\hline
Concept & Component & Numerical method / algorithm & Description\\
\hline
$\IP(\theta,m|D,M)$ & \code{Sampler} & \code{EMCEE}, \code{SABC}, \code{MCMC}, \code{MC} & parameters sampling\\
$L(D|\theta,M)$ & \code{Likelihood} & \code{Direct}, \code{PF} & states sampling and likelihood\\
$\rho(D|\theta,M)$ & \code{Distance} & \code{Norm}, \code{Regression}& distance for ABC sampler\\
$m_t \sim M_{t}(\theta)$ & \code{Model} & \code{Randomwalk}, \code{External}, \dots & model for user's application\\
$\Pi(\cdot),\Sigma(\cdot)$, etc. & \code{Aggregator} & \code{Trajectories}, \code{Replicates} & aggregator of components\\
\hline
\end{tabular}
\end{center}

\subsection{SPUX component assignments and built-in services}
\label{s:components}

\begin{figure}[!ht]
\centering
\includegraphics[width=\textwidth]{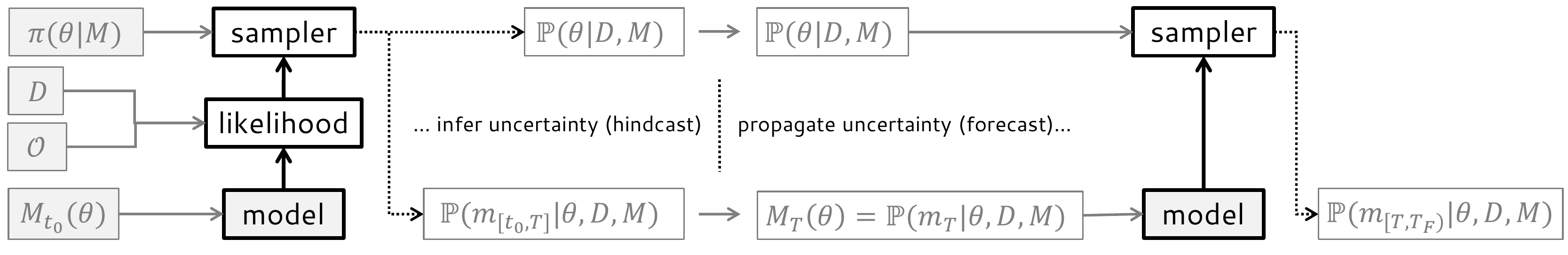}
\caption{Component assignment and execution flow scheme for the SPUX framework using
Markov-type sampler and likelihood components for the inferenece
(hindcast up to the last time $T$ in dataset $D$)
and the resulting bootstrapped posterior parameters and model states distributions
for the forecast (uncertainty propagation beyond the last time $T$ in dataset $D$).
Boxes with thin outlines indicate the associated mathematical objects
introduced in \autoref{s:concepts},
and boxes with thick outlines and thick arrows indicate SPUX components
and their internal assignments, respectively.
Thin solid and dotted arrows represent component inputs and outputs, respectively.
White background indicates built-in components and inference outputs,
whereas gray background indicates anticipated framework inputs.}
\label{f:spux-framework-summary-scheme}
\end{figure}

All available SPUX components can be assigned to each other following the required dependencies.
An example scheme for such assignment of the components required by any Markov-type sampling algorithm
for Bayesian inference is provided in \autoref{f:spux-framework-summary-scheme}
(the left part), together with the associated mathematical objects introduced in \autoref{s:concepts}.
The right part of this same scheme depicts an assignment of the components together with the associated mathematical objects
for the a posteriori forecast stage, introduced in \autoref{s:propagation-forecasting}.
Each SPUX component (for both stages: hindcast and forecast)
is described in detail in \autoref{s:usage}.
Such modularity in SPUX allows easy implementation of different numerical
approaches for Bayesian inference.
For instance, \autoref{s:rw-abc} explain how to use
a structurally different ABC-type method (see \autoref{s:sampling-abc}) within SPUX.

\subsection{SPUX parallelization capabilities}
\label{s:parallelization-scheme}

\begin{figure}[!ht]
\centering
\includegraphics[width=\textwidth]{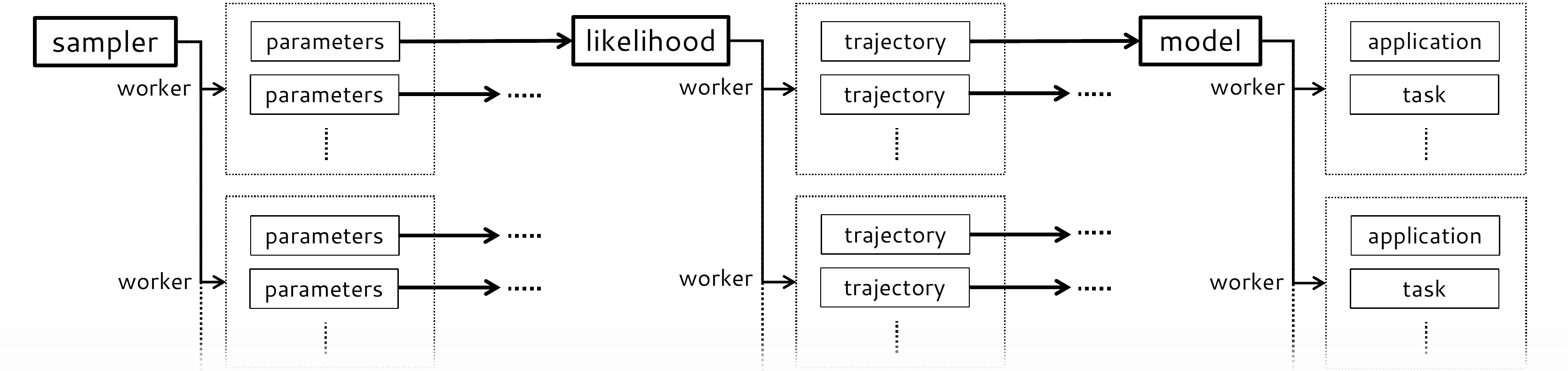}
\caption{An example of parallelization capabilities for various compontents
of the SPUX framework described in \autoref{s:components} and depicted in \autoref{f:spux-framework-scheme}.
Three levels of hierarchical parallelization are used here: parallelization
over multiple model parameters samples of the \code{EMCEE} sampler,
over multiple particles of the \code{PF} likelihood,
and over multiple independent tasks of a parallel user's application.
For each SPUX component (depicted using thick lines), the thin solid arrows represent
the required (possibly independent) tasks to be executed,
the thick solid arrows represent the internally called methods,
and the dotted lines represent the remaining ommited parts of the scheme.}
\label{f:spux-parallelization-scheme}
\end{figure}

One of the key advantage of SPUX is its very transparent yet very flexible parallelization sub-system.
In particular, multiple parallel workers can be attached to each spux component
listed in \autoref{s:components} and depicted in \autoref{f:spux-framework-summary-scheme}.
An example of such parallelization scheme with only three
components (sampler, likelihood, and model)
is provided in \autoref{f:spux-parallelization-scheme},
where sampler samples, likelihood particles,
and multiple tasks of the (optionally) parallel user's application/model
(more details in \autoref{s:parallel:model})
are distributed over available attached workers.
Additional parallel components can be incorporated when needed;
for instance, the \code{Replicates} aggregator is designed
to assimilate multiple independent observational data sources in parallel.
Note, that neither the model (i.e.~the associated user's application)
nor any other SPUX component is strictly required to be parallelized;
instead, any of these components might also be serial
(i.e. without any parallel workers attached).
For a more detailed description of parallel hierarchically stackable SPUX executors,
refer to the technical \autoref{s:parallelization}.

\subsection{SPUX gallery}
\label{s:gallery}

Inspired by the "Demo Data as Code" concept \cite{acm-data-as-code},
SPUX documentation website 
hosts a gallery listing examples
of user's applications, including source codes, authors, scientific fields, model programming languages,
used computational environment and configuration, figures with representative results,
and associated scientific publications.
At the time of this writing, example fields include
hydrology, aquatic ecology, urban hydrology, limnology, physics and data science,
but the generality of the SPUX framework does certainly extend beyond.
In particular, SPUX is currently being actively used for Bayesian inference in
realistic individual-based modelling of riverine macro invertebrates,
time-dependent conceptual hydrological modeling of catchments,
stochastic-input driven hydrological modeling of the rainfall runoff systems,
and high resolution three-dimensional hydrological (and, eventually, ecological) operational modeling of Lake Geneva.
%

\section{SPUX framework showcase with a random walk model}
\label{s:usage}

This section guides the reader through an example model and usage pattern of SPUX.
An overview of different SPUX installation methods can be found on the SPUX
documentation page, where we also provide the links to access the source code,
and a pre-configured SPUX Jupyter notebook (offered on a best-effort basis only).
After a brief overview of model types in \autoref{s:model:types},
an example model of a random walk and its setup in SPUX are described in \autoref{s:rw}.
The serial model execution procedure with a brief overview of resuts is
described in \autoref{s:rw-serial}.
In \autoref{s:rw-parallel}, minor auxiliary setup and execution steps
needed to run SPUX in parallel on workstations and high performance
clusters are addressed.
A detailed automatic PDF report compiling inference setup, results, diagnostics and performance
is introduced and interpreted in \autoref{s:rw-report}.
Finally, \autoref{s:rw-trajectory} and \autoref{s:rw-forecast-sequential}
describe procedures building upon already
estimated posteriors for model parameters and states:
re-executing the best (or any other) model trajectory
and forecasting to future times or performing sequential Bayesian updating for a new dataset.

\subsection{Deterministic and stochastic models}
\label{s:model:types}

SPUX supports all types of models for Bayesian inference introduced in \autoref{s:concepts}:
deterministic, where model evaluation is uniquely determined by parameters $\theta$,
and stochastic,
where model state depends on random variable(s) (e.g. initial data)
and/or is driven by stochastic process(es).

In Bayesian inference for deterministic models $M$,
a simple \code{Direct} likelihood can be analytically computed
using the specified error:
$L(D|\theta,M) = \Pobs (D|y,\theta)$ with $y = h(m,\theta)$ and $m = M(\theta)$.
An example of such model, \code{Straightwalk},
is available at \url{examples/straightwalk}.

For stochastic models $M$, in addition to uncertain model parameters $\theta$,
also the uncertain model states $m \sim M(\theta)$
need to be inferred.
In such cases, the error $\Pobs (D|y,\theta)$
by itself is often not sufficient to analytically compute the likelihood $L(D|\theta,M)$
in \eqref{eq:L-marginal}
for given model parameters. Hence, additional approximation techniques are
required, as discussed in \autoref{s:concepts}.
An example of a stochastic time-independent model (left part of \autoref{f:spux-framework-summary-scheme})
is provided in \code{examples/gaussian-sabc}.
Another built-in model in SPUX is a stochastic version of the \code{Straightwalk}
model, called \code{Randomwalk}, where the direction and size of each (time) step
of the walker is a random variable.
Since the SPUX framework is tailored to Bayesian inference
with stochastic time-dependent models,
in the following section we showcase
the \code{Randomwalk} model.
Files mentioned throughout next section are located
in SPUX repository at \url{examples/randomwalk}
(assumed to be the current working directory).

\subsection{Randomwalk model}
\label{s:rw}

The \code{Randomwalk} model describes a stochastic walk on a line (i.e.~a set of real numbers).
Its goal is to provide the simplest possible conceptual model,
which has just enough complexity to illustrate most of the functionality in SPUX,
while addressing the majority of requirements in the environmental scientific modeling.
Given a prescribed time step size $\Delta t$ (in seconds for example, to fix the ideas)
as a numerical discretization parameter,
together with an initial time $t_0$ and an initial (possibly uncertain) position $m_{t_0} \sim M_{t_0}$ [m],
the model iteratively takes random steps on a one-dimensional line
every $\Delta t$ time units. The direction and the size of each step depend on the time
step size, on two models parameters, the drift $\mu$ [m/s] and the volatility $\sigma$ [$ms^{-1/2}$],
and on an independent standard normal random variable $\cN_t$:
\begin{equation}
\label{eq:rw}
m_{t + \Delta t} \sim \IP(m_{t + \Delta t}|m_t) = m_t + \Delta t \mu + \sqrt{\Delta t} \sigma \cN_t.
\end{equation}

The \code{Randomwalk} model is a built-in SPUX model class
with model initialization and evolution (according to \eqref{eq:rw})
implemented in the corresponding class methods:
\begin{center}
\begin{tabular}{l|l}
\hline
\code{init(...)} & initialize model with \code{initial} state $m_{t_0}$ and model \code{parameters} $\theta$\\
\code{run(...)}  & run model up to the specified \code{time} $t$ and return model prediction output $y_t$\\
\hline
\end{tabular}
\end{center}
For implementation details of this demonstrational model,
focused on simplicity and code readability (no vectorization, explicit \code{for}-loops),
refer to \autoref{lst:randomwalk} in \autoref{s:model-scripts}.
In this particular model, the \code{initial} argument contains the initial time $t_0$ and the initial position $m_{t_0}$.
Alternatively, \code{initial} contents might as well be assigned
in the model constructor directly, as for the time step size $\Delta t$.
However, such explicit specification of the model input (i.e., using \code{initial}) at the initialization
provides more flexibility in the cases where the initial position $M_{t_0}$ is uncertain (see \autoref{s:rw-serial})
and/or when multiple observational datasets are available (see \autoref{s:replicates}).
Finally, we assume $h (m,\theta) = m$, i.e.~model state coincides with the model output.

\subsection{Inference for Randomwalk model}
\label{s:rw-serial}

\begin{figure}[!ht]
\centering
\includegraphics[width=\textwidth]{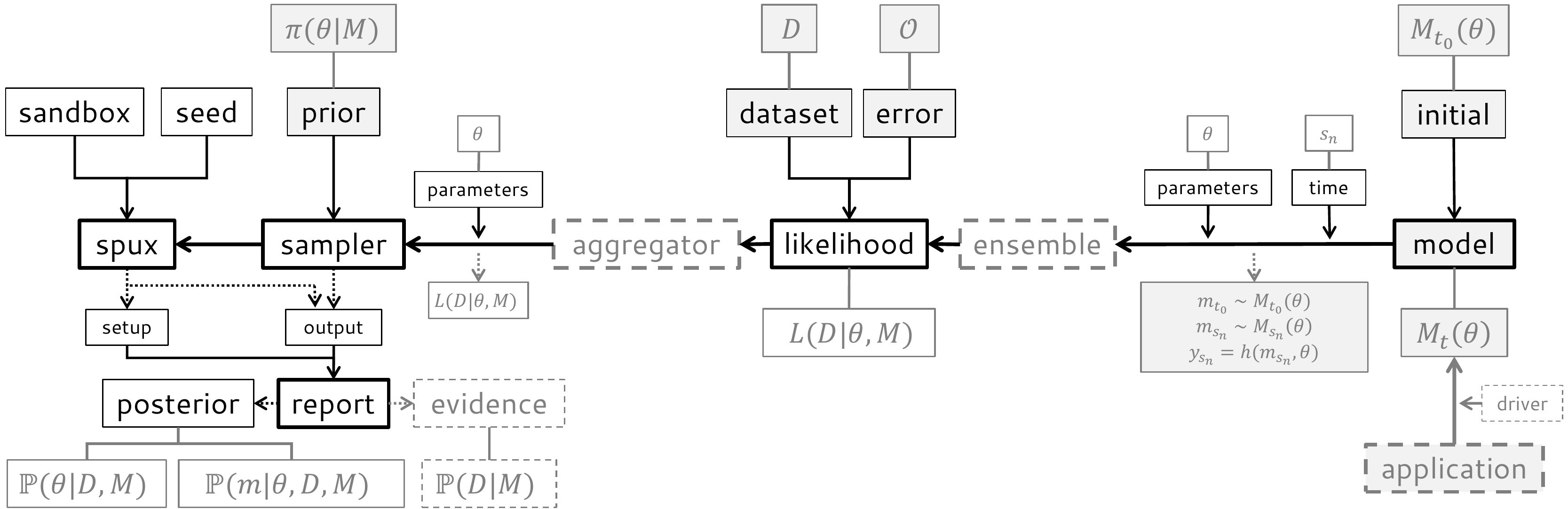}
\caption{
Component assignment and execution flow scheme for the SPUX framework using
Markov-type sampler and likelihood components
(an extension of the summary \autoref{f:spux-framework-summary-scheme}).
Boxes with thin outlines indicate the associated mathematical objects
introduced in \autoref{s:concepts},
and boxes with thick outlines and thick arrows indicate SPUX components
and their assignments for internal evaluation, respectively.
For each SPUX component,
thin solid and dotted arrows represent component inputs
and outputs, repspectively.
White background indicates built-in components and inference outputs,
whereas gray background indicates anticipated framework inputs.
Dashed boxes indicate optional SPUX components for special
sampler, likelihood, model or dataset requirements
and optional output or reporting capabilities.}
\label{f:spux-framework-scheme}
\end{figure}

As briefly described in \autoref{s:components},
all available SPUX components can be assigned among each other following the required
dependencies. An extended version of the assignment scheme in \autoref{f:spux-framework-summary-scheme}
for this example,
together with the associated mathematical objects introduced in \autoref{s:concepts},
is provided in \autoref{f:spux-framework-scheme}. The final assigment assigns the top component,
in this case (and also usually) the sampler, to the built-in main SPUX framework component.
The SPUX framework component provides
the hierarchical sandboxing (isolation to dedicated directories)
and seeding (controlling the independence of random number streams and ensuring
reproducibility independently of the chosen computational environment),
manages runtime checkpointing, diagnostics,
and framework setup options (see later \autoref{s:framework-setup}).
In the following subsections we provide an elaborate description
of the remaining SPUX framework prerequisites, supplementary capabilities,
and a general inference execution workflow,
using the \code{Randomwalk} model (introduced in \autoref{s:rw}) as the underlying example.

\subsubsection{SPUX configuration prerequisites}
\label{s:usage-prerequisites}

To perform Bayesian inference for the given model,
we also need to specify several configuration files for
the remaining essential prerequisites (besides the \code{sampler}, the \code{likelihood}
and the \code{model}).
In particular, for the \code{Randomwalk} example, these prerequisites are: the available observational data $D$,
a statistical error $\Pobs (D|M(\theta),\theta)$,
prior distributions $\pi (\theta|M)$ for all the parameters
that we intend to infer, and the initial model state $M_{t_0}(\theta)$.
Note, that the parameters $\theta$ contain both the parameters of the model $M$
and, if present, the hyper-parameters used for constructing the distributions of
the error $\Pobs (\cdot|y_s,\theta)$ and/or of the initial model states $M_{t_0} (\theta)$.
Prior initial model state distribution $M_{t_0}(\theta)$ is then attached
to the specified \code{model} component (representing $M(\theta)$),
error $\Pobs (D|M(\theta),\theta)$ is attached to \code{likelihood} component (representing $L(D|\theta,M)$),
and prior model parameters distribution $\pi (\theta|M)$ is attached to \code{sampler} component (representing $\IP(\theta,m|D,M)$),
see \autoref{s:components} and \autoref{f:spux-framework-scheme}.

Marginal prior distributions $\pi_{\mu}$, $\pi_{\sigma}$, $\pi_{\varepsilon}$
for each parameter $\theta = (\mu, \sigma, \varepsilon)$
of the joint prior distribution $\pi(\theta|M)$,
with $\varepsilon$ being the standard deviation (as a hyper-parameter)
of the (assumed) Gaussian error (described below),
are assumed to be independent and are given by
(for plots, refer to \autoref{f:randomwalk_parameters}):
\begin{equation}
\label{eq:rw-prior}
\pi_{\mu} = \cU (-1, 1) \ [m/s], \quad
\pi_{\sigma} = \cU (5, 15) \ [ms^{-1/2}], \quad
\pi_{\varepsilon} = \cL\cN (1, 1) \ [m].
\end{equation}
This multivariate prior distribution is defined in \url{prior.py}
using a built-in \code{Tensor} SPUX distribution
(see \autoref{s:add:distribution} for an overview of all SPUX distributions).
The \code{Tensor} combines multiple statistically independent distributions
(provided as a Python dictionary of the corresponding univariate distributions
- for instance, from the \url{scipy.stats} package)
to a multivariate SPUX distribution.
The initial position $m_{t_0}$ of the randomwalk is also uncertain,
and hence a prior distribution $M_{t_0}$ for the initial position $m_{t_0}$ at (deterministic) time $t_0 = 0$
is defined in \url{initial.py}:

%
\begin{equation}
\label{eq:rw-initial}
m_{t_0} \sim M_{t_0} (\theta) = \cN (10, 2) \ [m], \quad t_0 \sim \delta_{0} \ [s].
\end{equation}
We note, that $m_{t_0}$ could alternatively be included as a model parameter
in the prior defined in \eqref{eq:rw-prior}.

Actual dataset files are located in the \url{datasets} directory.
The default container for dataset management in SPUX is
a \code{DataFrame} of the \code{pandas} package \cite{pandas},
which is very similar to the dataframe in the \code{R} programming language.
The example script to load the dataset into a \code{DataFrame},
is located in \url{dataset.py}.
The dataset provides \emph{inaccurate} observations $D_s$
of the position $m_s$ at (snapshot) times $s = s_1, \dots, s_N$.
N/A values are allowed, however,
no column (quantity of interest) or row (snapshot)
should contain \emph{only} N/A values.

The inaccuracies in observational data are modeled with an error,
which is a statistical distribution of the observations (from the dataset)
conditional on the specified model output prediction (from the simulation).
In this example this distribution is assumed to be normally distributed with mean
equal to the position $m_s$ predicted by the model, and with standard deviation
given by an additional uncertain parameter $\varepsilon$ used as a hyper-parameter:
\begin{equation}
\label{eq:rw-error}
D_s \sim \cO (\cdot | y_s, \theta) = \cN (y_s, \varepsilon) \ [m].
\end{equation}
All observations are assumed to be statistically independent.
The observational error is defined in \url{error.py}
as a function (or a collable object)
which, given the model output prediction and parameters,
constructs the statistical distribution above as
a SPUX \code{Distribution} instance.

An (optional) dictionary specifying units (as LaTeX-supported strings)
for each parameter, model output and time dimension is specified in \code{units.py}.
Its contents are used only in the SPUX report.

Within the context of this illustrative example,
we also make use of the (optional) exact (loaded in \url{exact.py})
parameter values available at \url{datasets/exact.dat}
and the exact (synthetic) model outputs (without the observational noise)
available at \url{datasets/predictions.dat}.

\subsubsection{SPUX configuration and execution - user interface (UI)}
\label{s:usage-conf-file}

The quickest way to run SPUX inference and post-processing (discussed in the following sections)
is using the SPUX UI configuration file \url{spux.cfg}.
There, arguments required for the construction, configuration,
initialization, and execution of each SPUX component (either built-in or manually imported) can be specified,
including any prerequisite script (see \autoref{s:usage-prerequisites}) defining required (mandatory or optional) options.
All such components and options can be configured as depicted in \autoref{lst:spux.cfg},
where an adaptive Particle Filter (\code{PF} with the default "rejection" smoothing) likelihood
with the specified maximum number of particles
is assigned to an Affine Invariant Ensemble sampler (\code{EMCEE})
with the specified number of concurrent chains.
Additional framework options (i.e.~not specific to any component)
can also be specified in \url{spux.cfg}
to control various aspects of the SPUX framework, as depicted in \autoref{lst:spux-options.cfg}.
For instance, optional \code{units} of time, model parameters and observations
can be specified (see also \autoref{s:add:distribution}).
Built-in component and framework options
are described throughout the following sections
with a summary available in \autoref{s:framework-setup}.

\noindent
\begin{minipage}[t]{.47\textwidth}
\begin{lstlisting}[caption={Example \url{spux.cfg} for components},label={lst:spux.cfg},style=python]
model Randomwalk
model.dt 0.1
likelihood PF
likelihood.particles 256
sampler EMCEE
sampler.chains 32
sampler.samples 10000
\end{lstlisting}
\vfill
\end{minipage}
\hfill
\begin{minipage}[t]{.47\textwidth}
\begin{lstlisting}[caption={Additional options in \url{spux.cfg}},label={lst:spux-options.cfg},style=python]
prior prior.py
error error.py
dataset dataset.py
initial initial.py
burnin "half"
units units.py
exact exact.py
\end{lstlisting}
\end{minipage}
\\

\noindent
Using SPUX UI, inference and post-processing can be setup and performed
by simply executing:
\begin{center}
\code{spux spux.cfg {-}{-}execute {-}{-}all}
\end{center}
This automatically generates the required SPUX scripts
(described in the following sections)
and automatically executes testing, synthesis, inference, reporting,
and re-execution of the best trajectory.

The inference process can be terminated at any time,
since the output is periodically checkpointed (see \autoref{s:framework-setup}).
Additional runtime arguments can be specified to customize the inference:
\begin{center}
\begin{tabular}{r|l}
\hline
\code{{-}{-}dry} & "dry run" mode - inspect configuration without actual sampling\\
\code{{-}{-}continue} & continue the inference process starting from the latest checkpoint\\
\code{{-}{-}no-repro} & disable reproducibility information (stored in \url{randomwalk_reprozip.rpz})\\
\hline
\end{tabular}
\end{center}
The list of all built-in components and options
is also retrievable by executing \code{spux {-}{-}help}.

\subsubsection{SPUX configuration and execution - application programming interface (API)}
\label{s:usage-conf-script}

The most flexible way to configure SPUX is to use a configuration script in which
the prior, error model, and dataset are explicitely
imported and assigned (using the SPUX API)
to the selected SPUX components, such as likelihood (or distance) and sampler.
A summary of an example \url{configure.py} (excluding trivial module imports)
is provided in \autoref{lst:configure}.
The mandatory \code{spux.assign(...)}
assigns all hierarchically ordered components to the SPUX framework.

\noindent
\begin{minipage}[t]{.61\textwidth}
\begin{lstlisting}[caption={Example \url{configure.py} (excluding imports)},label={lst:configure},style=python]
model = Randomwalk (dt=0.1)
model.configure (initial)
likelihood = PF (particles=256)
likelihood.configure (dataset, error)
sampler = EMCEE (chains=32)
samper.configure (prior, samples=10000)
spux.configure (units)
spux.assign (model, likelihood, sampler)
\end{lstlisting}
\vfill
\end{minipage}
\hfill
\begin{minipage}[t]{.36\textwidth}
\begin{lstlisting}[caption={Example \url{infer.py}},label={lst:infer},style=python]
from configure import *
spux.setup ()
spux.init ()
sampler.init ()
sampler ()
spux.exit ()
\end{lstlisting}
\end{minipage}
\\

The execution script \code{infer.py} provided in \autoref{lst:infer}
imports the components from \code{configure.py},
setups the SPUX framework (mandatory \code{spux.setup(...)}, see \autoref{s:framework-setup}),
initializes the sampler (mandatory \code{sampler.init(...)} for EMCEE),
and performs the posterior sampling of the model parameters and states
for the specified number of samples.
The mandatory framework initialization \code{spux.init(...)} and finalization \code{spux.exit(...)}
methods manage the required computational resources.
%
%
For the EMCEE sampler,
initial model parameters are drawn by default from the specified prior.
The script can be executed by typing \code{python infer.py} in the console.
Analogously to the UI in \autoref{s:usage-conf-file},
the \code{{-}{-}dry}, \code{{-}{-}continue} and \code{{-}{-}no-repro}
runtime arguments can be used for \code{infer.py}
to enable "dry mode", continuation or disable reproducibility package.
%

\subsubsection{SPUX results}
\label{s:usage-results}

\begin{figure}[!ht]
	\begin{center}
		\includegraphics[width=1.00\textwidth]{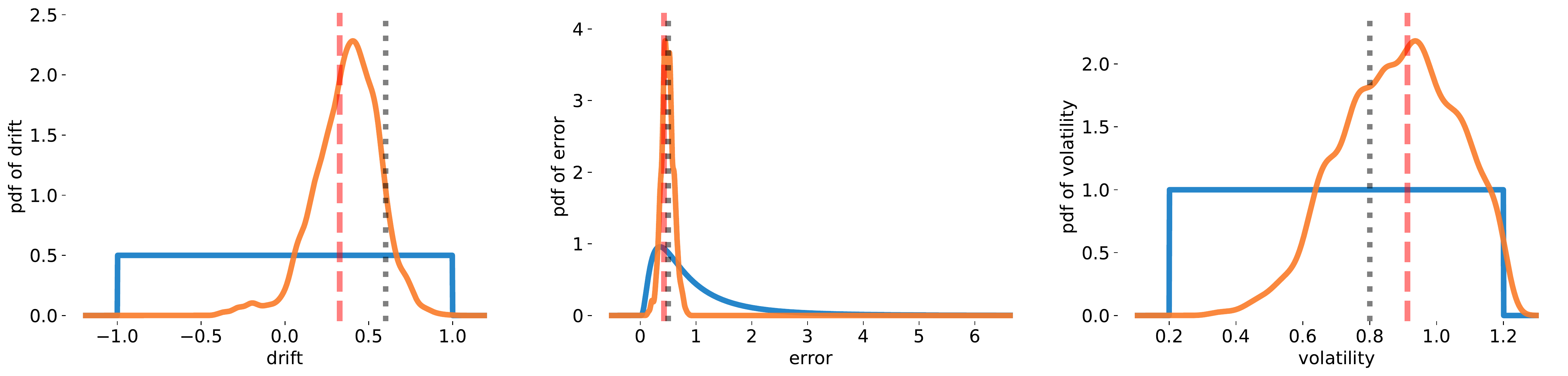}
	\end{center}
    \caption{Marginal posterior (orange) and prior (blue) distributions of model parameters.
The red dashed line indicates the best found parameters values.
The black dotted line represents the exact parameter values.}
	\label{f:randomwalk_parameters}
\end{figure}

\begin{figure}[!ht]
	\begin{center}
		\includegraphics[width=0.96\textwidth]{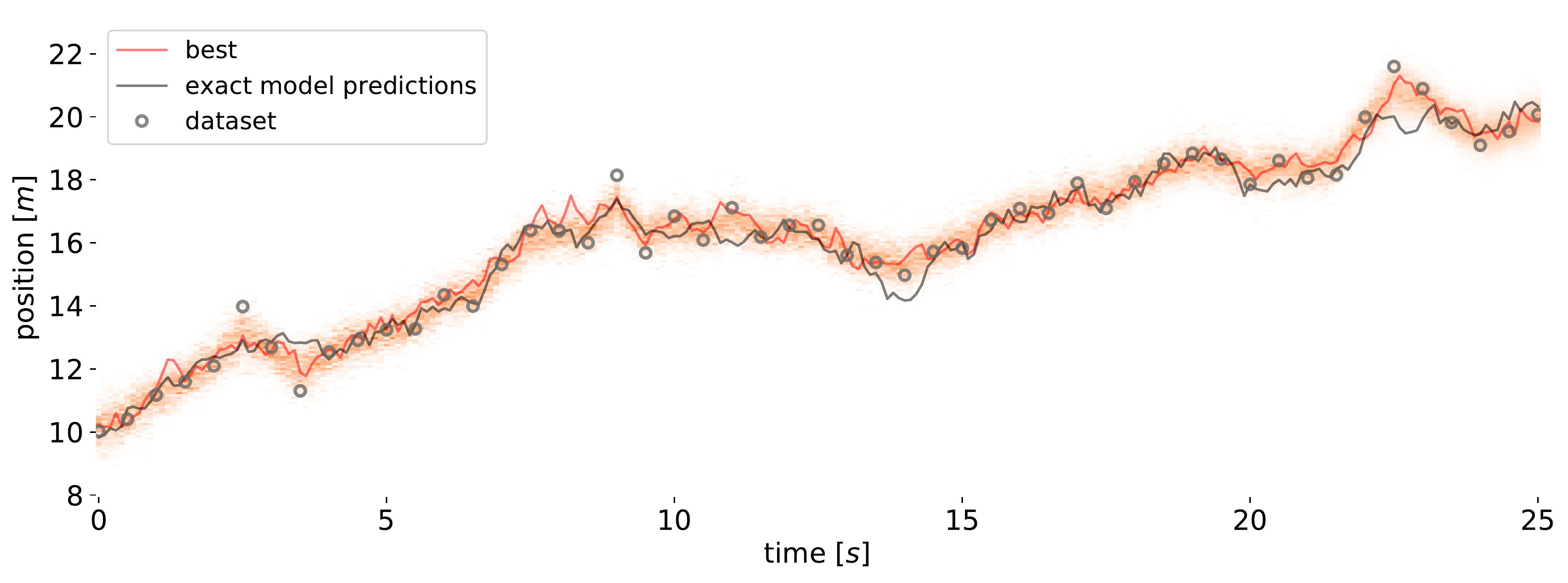}
	\end{center}
    \caption{Posterior distribution of model predictions for the observational dataset.
The shaded orange regions indicate
the log-density of the posterior model predictions distribution at the respective time points,
the red line represents the best found model prediction,
the black line represents the exact model prediction values.}
	\label{f:randomwalk_predictions}
\end{figure}

The estimated marginal posteriors of model parameters are provided in \autoref{f:randomwalk_parameters}
and the estimated marginal posteriors of model predictions are provided in \autoref{f:randomwalk_predictions}.
%
%

For the inference results,
the default burnin period (half of all samples) was selected
to remove the initial sampler bias.
The adaptive number of particles
described later in \autoref{s:pf-adaptivity}
is locked after the specified \code{lock} batches.
By default the burnin is also set to this value
to avoid any potential bias due to the adaptivity process.
For post-processing,
only every \code{thin}-th sample (of each sampler chain) is selected
in order to obtain a sequence of statistically independent posterior samples.
In particular, the default \code{"auto"} value was used for the \code{thin} period,
which uses the median of optimal thinning periods
obtained by estimating multivariate effective posterior sample sizes for each sampler chain.

\subsection{Parallel inference for Randomwalk model}
\label{s:rw-parallel}

With minimal effort, the above example configuration can be parallelized
either for a local machine or for a remote high performance computing (HPC) cluster.
We emphasize, that no modifications are needed for this particular "Randomwalk" model class.
For HPC cluster,
consider placing (see \autoref{s:framework-setup}) the \code{output} directory
in a parallel high performance "scratch" filesystem, if available.
For a more detailed discussion regarding models not written in pure Python,
refer to \autoref{s:customization}.

\subsubsection{Attaching parallel workers}
\label{s:workers}

To enable parallel execution, a required number of parallel workers can be attached
to each SPUX component, as depicted in \autoref{f:spux-framework-summary-scheme}.
Examples are provided in
\autoref{lst:spux-workers.cfg} (for the UI configuration file \code{spux.cfg} as in \autoref{lst:spux.cfg})
and in
\autoref{lst:infer-workers} (for the API inference script \code{infer.py} as in \autoref{lst:infer}, in
this case these lines should be placed before the calls to framework setup and initialization).

\noindent
\begin{minipage}[t]{.43\textwidth}
\begin{lstlisting}[caption={Parallel workers in \code{spux.cfg}.},label={lst:spux-workers.cfg},style=python]
likelihood.workers 8
sampler.workers 16
\end{lstlisting}
\vfill
\end{minipage}
\hfill
\begin{minipage}[t]{.51\textwidth}
\begin{lstlisting}[caption={Parallel workers in \code{infer.py}.},label={lst:infer-workers},style=python]
likelihood.attach (workers=8)
sampler.attach (workers=16)
\end{lstlisting}
\end{minipage}
\\

A separate dedicated core is used for the manager process of each group of parallel workers.
For an advice regarding worker allocation strategies across multiple parallel executors,
refer to \autoref{s:remark-workers}.
Advanced explicit attaching of SPUX executors to components is described in \autoref{s:executors-types}.

\subsubsection{Launching parallel SPUX}
\label{s:launch}

Assuming a library for the Message Passing Interface (MPI) \cite{MPI3} is installed,
parallel scripts need to be launched through the Python \code{mpi4py} module.
For the execution using the UI, specify \code{{-}{-}mpi} runtime argument.
For \code{infer.py} using the API:
\begin{center}
\code{mpiexec -n 1 python -m mpi4py infer.py}.
\end{center}
The required worker MPI processes
will be spawned automatically (i.e.~according to the resources table).

For HPC systems not supporting dynamical spawning of new MPI processes,
the required number of MPI ranks (workers) needs to be explicitly specified for \code{mpiexec} (after "\code{-n}").
The cumulative number of required workers is indicated in the bottom right cell
of the computational resources table (see example in \autoref{R-t:randomwalk_environment}),
which is printed to the console
already during the "dry run" mode (for which one core is sufficient, i.e.~no MPI is required).
For convenience (e.g.~to automate parallel job submission process),
this number is also written to the dedicated \code{workers.txt} file.
Parallelization can be temporarily disabled with \code{{-}{-}serial} runtime argument,
which ignores all workers attachments.
SPUX documentation outlines specifics regarding different MPI libraries
and useful advice to address any potential issues.

\subsection{SPUX report}
\label{s:rw-report}

All tables and figures generated by the SPUX framework,
such as previous \autoref{f:randomwalk_parameters} and \autoref{f:randomwalk_predictions},
are automatically included (see \autoref{s:usage-setup-output} for technical details)
in a PDF report (A4 and "slides" layouts), described in later sections.
Such PDF report is also provided to support this section on SPUX usage as Suplementary Material.
All of the tables, figures, and even the configuration scripts (in \autoref{R-s:report-scripts}) referenced within this section are available
in this report, hence it is strongly advisable to have a separate copy of the Suplementary Material at hand.
Additionally, the "reproducible package" \url{spux.rpz} is generated using the "reprozip" tool \cite{ChirigatiRSF16}.
The SPUX report and the \url{spux.rpz} provide the highest level of reproducibility (excluding containerization
techniques) for an inference or forecast run, independently of the chosen computational environment and/or
hardware.

\subsubsection{Configuration and setup section}
\label{s:report-conf}

SPUX configuration and setup is summarized
in the first section of the SPUX report (see \autoref{R-s:report-setup}),
which is generated by executing the example \url{report.py} script and contains the following:
\begin{center}
\begin{tabular}{r|c|l}
\hline
configuration & \autoref{R-t:randomwalk_configuration} & SPUX configuration: component classes and their options\\
setup & \autoref{R-t:randomwalk_setup} & framework options from \code{spux.setup(...)}, see \autoref{s:framework-setup}\\
units & \autoref{R-t:randomwalk_units} & units for parameters, observations, and time\\
exact & \autoref{R-t:randomwalk_exact} & exact model parameters (if specified)\\
evaluations & \autoref{R-t:randomwalk_evaluations} & total number of anticipated model evaluations across components\\
\hline
datasets & \autoref{R-f:randomwalk_datasets} & dataset(s) $D$ and exact model predictions (if specified)\\
errors & \autoref{R-f:randomwalk_errors} & marginal error models $\Pobs$ distributions for the specified $\theta, y$\\
prior & \autoref{R-f:randomwalk_prior} & marginal prior distributions of model parameters - $\pi(\theta,M)$ \\
initials & \autoref{R-f:randomwalk_initials} & marginal prior distributions of initial model states - $M_{t_0} (\theta)$\\
\hline
\end{tabular}
\end{center}
In particular, for the \code{Randomwalk} example,
at most 256 particles were used in the PF likelihood (with adaptivity enabled, see \autoref{s:pf-adaptivity}),
and 32 chains were used in the EMCEE sampler.
In total, 10'000 samples were requested, locking particle adaptivity after 75 sample batches (2'400 samples).

\subsubsection{Results and diagnostics sections}
\label{s:report-results}

To load and visualize inference results and diagnostics using the API,
the \url{report.py} script is executed with the additional \code{{-}{-}results} runtime argument.
This \code{report.py} script uses built-in plotting routines available in \code{spux.reports.mpl} module.
The user can freely choose to use the reconstructed results and diagnostics
with other established data visualization libraries, including the specialized
\code{pandas.plotting} module and \code{arviz} \cite{arviz_2019} package.
The report script generates multiple tables and figures of the results and diagnostics,
and updates the SPUX report accordingly.

The inference results tables and figures included in \autoref{R-s:report-results} of the SPUX report
provide posterior distributions for model parameters and predictions (i.e.~model outputs $y$ or even model states $m$):
\begin{center}
\begin{tabular}{r|c|l}
\hline
best-parameters & \autoref{R-t:randomwalk_best-parameters} & best found (e.g.~maximum a posteriori) model parameters\\
\hline
parameters & \autoref{R-f:randomwalk_parameters} & marginal posteriors $\IP(\theta|D,M)$ of model parameters $\theta$\\
predictions & \autoref{R-f:randomwalk_predictions} & marginal posteriors $\IP(m|D,M)$ of model state $m$\\
dependencies2d & \autoref{R-f:randomwalk_dependencies2d} & dependencies among pairs of posterior parameters $\theta$ or states $m$\\
parameters2d & \autoref{R-f:randomwalk_parameter2d-strongest} & joint posterior for the most dependent model parameters $\theta$ pair\\
\hline
\end{tabular}
\end{center}
In particular \autoref{R-f:randomwalk_parameters} and \autoref{R-f:randomwalk_predictions},
are included here as \autoref{f:randomwalk_parameters} and \autoref{f:randomwalk_predictions},
respectively.

Additional "diagnostics" tables and figures are included
in \autoref{R-s:report-diagnostics} of the SPUX report,
providing
quality assesments of the inference results
and the algorithmic technicalities for
the Markov chain sampling (\code{EMCEE} in this case),
as well as the likelihood estimation (\code{PF} in this case):
\begin{center}
\begin{tabular}{r|c|l}
\hline
status & \autoref{R-t:randomwalk_status} & information about loaded SPUX \code{status}, see \autoref{s:usage-setup-output} \\
metrics & \autoref{R-t:randomwalk_metrics} & metrics such as effective sample size, thinning period, etc. \\
\hline
residuals & \autoref{R-f:randomwalk_residuals} & residuals (differences between the dataset and outputs)\\
QQ & \autoref{R-f:randomwalk_qq} & quantile-quantile comparison of residuals and $\Pobs$ distributions\\
successfuls & \autoref{R-f:randomwalk_successfuls} & tracking of the failed or skipped likelihood evaluations\\
samples & \autoref{R-f:randomwalk_samples} & progress of model parameters sampling (including burnin)\\
samples-cutoff & \autoref{R-f:randomwalk_samples-cutoff} & progress of model parameters sampling (excluding burnin)\\
acceptances & \autoref{R-f:randomwalk_acceptances} & progress of the instantaneous sampler acceptance rate\\
resets & \autoref{R-f:randomwalk_resets} & tracking likelihood re-estimations due to stuck chains\\
autocorrelations & \autoref{R-f:randomwalk_autocorrelations} & autocorrelations of Markov chain parameters samples\\
likelihoods & \autoref{R-f:randomwalk_likelihoods} & progress of prior/likelihood/posterior (including burnin)\\
likelihoods-cutoff & \autoref{R-f:randomwalk_likelihoods-cutoff} & progress of prior/likelihood/posterior (excluding burnin)\\
fitscores & \autoref{R-f:randomwalk_fitscores} & progress of fitscores as in \autoref{s:pf-adaptivity-fitscores} (including burnin)\\
accuracies & \autoref{R-f:randomwalk_accuracies} & progress of likelihood accuracy as in \autoref{s:pf-adaptivity-accuracies}\\
particles & \autoref{R-f:randomwalk_particles} & progress of likelihood adaptivity as in \autoref{s:pf-adaptivity-procedure}\\
redraw & \autoref{R-f:randomwalk_redraw} & progress of the particle redraw fraction in \code{PF} (see \autoref{f:spux-sampler-pf})\\
redraw-temporal & \autoref{R-f:randomwalk_redraw-temporal} & temporal progress of the redraw fraction in \code{PF} (see \autoref{f:spux-sampler-pf})\\
\hline
\end{tabular}
\end{center}

From these diagnostic plots,
also included in \autoref{R-s:report-diagnostics} of the SPUX report,
we determine that the inference was relatively successful.
In particular, the effective sample size (computed using the estimated autocorrelations
as in \autoref{R-f:randomwalk_autocorrelations})
is not much smaller than the actual request by the sampler,
the posterior residuals distribution is consistent with the theoretical distribution
prescribed in the error model, not many failed (NaN - where model did not return an output)
or skipped (due to at least one proposed parameter laying outside the support of its prior)
likelihood evaluations, and converged sampling of the parameters space due to the stationarity of the Markov
chains.
Additionally,
the average acceptance rate is relatively satisfactory (considering there were 3 model
parameters), total chain resets (likelihood re-estimations) due to stuck
chains are negligible, and chain autocorrelations lengths are relatively short.
The adaptivity within the \code{PF} (described in \autoref{s:pf-adaptivity})
is also successful: fitscores below the prescribed threshold,
accuracies in the prescribed interval,
the number of particles steadily adapted within the specified limits during the burnin stage,
and the average redraw rate (the fraction of unique particles in particle filter after each resampling)
well above half the total number of particles (indicating the absence of any critical degeneration,
e.g., collapsing on a single particle, of the \code{PF} resampling procedure).

Finally, various (approximate) criterions
for model suitability \cite{doi:10.1002/2017WR021902}
are provided in \autoref{R-t:randomwalk_metrics}:
\begin{center}
\begin{tabular}{r|l}
\hline
with $\Pobs$ &
Bayesian Model Evidence (BME),
Kashyap/Baysian Information Criterions (KIC/BIC)\\
w/o $\Pobs$ &
Bayesian Cross Validation (BCV),
Deviance/Akaike Information Criterions (DIC/AIC)\\
\hline
\end{tabular}
\end{center}
%
%
Bayesian factors
$K(M_a, M_b) = \IP(D|M_a) / \IP(D|M_b)$
for models $M_{a/b}$ from above metrics,
determine if model $M_a$ (relative to model $M_b$) is strongly supported ($K > 10$)
by the observational dataset(s).

\subsubsection{Computational environment and performance sections}
\label{s:performance-serial}

In \autoref{R-s:report-environment} of the SPUX report,
the computational environment and
attached computational resources
are provided:
\begin{center}
\begin{tabular}{r|c|l}
\hline
environment & \autoref{R-t:randomwalk_environment} & computational environment (date, time, hardware, software versions)\\
resources & \autoref{R-t:randomwalk_resources} & required computational resources in terms of workers (e.g.~cores)\\
\hline
\end{tabular}
\end{center}
In particular, for the \code{Randomwalk} example, we used 145 cores
in total, with 16 parallel workers for the \code{EMCEE} sampler,
and 8 parallel workers for the \code{PF} likelihood.

In \autoref{R-s:report-performance}, tables with
measured runtimes of the entire inference run are included:
\begin{center}
\begin{tabular}{r|c|l}
\hline
runtimes & \autoref{R-t:randomwalk_runtimes} & total inference runtimes (wall-clock and serial equivalent)\\
runtimes-latest & \autoref{R-t:randomwalk_runtimes-latest} & latest inference runtimes (wall-clock and serial equivalent)\\
\hline
\end{tabular}
\end{center}

Optional computational performance plots for \autoref{R-s:report-performance} of the SPUX report,
providing additional insight into
the computational and algorithmical efficiency of the inference process,
can be generated by specifying
\code{{-}{-}performance} runtime argument for the \code{report.py} script.
In particular, "runtimes" of key SPUX routines
are measured by default (see "performance" keyword for "informative" option in \autoref{s:framework-setup}).
This allows to generate the "runtimes" plot for the entire sampling progress
or more easily interpretable "runtime" plots for specific sampler batches.
%
Optionally, if "timestamps" keyword is requested for "informative" option,
the respective "timestamps" plots can be generated,
providing an insight into the detailed SPUX performance profiles.
Additionally, plots for parallel efficiencies for the entire sampling progress
and strong scaling (multiple SPUX executions using different number of parallel workers)
can be generated (refer to the SPUX documentation).
%
%
Since the current \code{Randomwalk} example takes virtually no time to be executed,
it is 
of little value
to investigate the computational performance of SPUX
here;
such detailed investigations will be included in the subsequent publications focused
on the application of the SPUX framework to realistic models and datasets.

%
%
%
It is, however, worthwhile to inspect the parallelization performance in terms of
the measured "traffic" types and amounts within the adaptive \code{PF} resampling process,
available in:
\begin{center}
\begin{tabular}{r|c|l}
\hline
traffic & \autoref{R-f:randomwalk_traffic} & progress of copied/moved particles and communication cost\\
traffic-temporal & \autoref{R-f:randomwalk_traffic-temporal} & temporal progress of copied/moved particles and comm. cost\\
\hline
\end{tabular}
\end{center}
Note, that due to the communication-avoiding load balancing in the resampling parallelization routing
(see \autoref{s:executors-balancing-ensemble}),
only a small fraction of all copied particles needs to be moved,
and the associated communication costs are even lower
due to the exploitation of the node-level affinity of the parallel workers
(to optimize processing of the "statefiles", see \autoref{s:add:serialization-sandboxing}).
The initial period is dominated by the "move" traffic,
since the number of parallel workers equals the initial number of particles,
allowing only the exploitation of the node-level affinity (if "statefiles" are used).
%

\subsection{Executing a selected (e.g. "best") model trajectory}
\label{s:rw-trajectory}

If required, the best (or any other) model trajectory,
corresponding to the best model parameters,
\emph{and} the best model predictions (which, for stochastic models, are not determined only by the best model parameters),
can be explicitly executed using the auto-generated \code{best.py} script.
Such a-posteriori explicit execution of a specified model trajectory
allows to configure the model for richer output,
that is otherwise not required during the inference (i.e.~for comparison with the datasets)
or not accessible using the functionality of the \code{history} option (see \autoref{s:framework-setup}).
For instance, instead of only the model output $y$, more of the hidden model state $m$ could be returned as model predictions.
Additionally, an explicit \code{trajectory} directory
is used for the model's sandbox (see \autoref{s:sandboxing})
instead of potentially inaccessible node-local filesystems (see \autoref{s:add:sandboxing}).

\subsection{Forecast (uncertainty propagation) and sequential Bayesian updating}
\label{s:rw-forecast-sequential}

In many use cases, datasets could be structured into multiple time periods,
allowing to perform the Bayesian inference sequentially,
and providing an optional future forecasts in-between such datasets.

Firstly, a forecast of the model predictions can be obtained 
by simply propagating the inferred posterior distributions from a preceeding 
time period $[t_0, T]$, for which a dataset of observed data is available,
to a future time period $[T, T_F]$, 
see \autoref{s:propagation-forecasting} and \autoref{f:spux-framework-summary-scheme}.
This can be achieved by specifying, within the UI configuration file of the SPUX framework,
the location of the "past" inference (with \code{states=True}, see \autoref{s:framework-setup}) root directory as \code{pastdir} and the list of "future" times as \code{timeset}
(see the example provided in \autoref{s:rw-forecast-assimilate-scripts}, \autoref{lst:spux-forecast.cfg}).
The UI automatically generates the corresponding \code{prior.py} and \code{initial.py} scripts
for the bootstrap distribution of samples from posterior model parameters and
the associated bootstrap distribution of samples from posterior model predictions
at the last snapshot of the dataset, respectively.
These prior and initial prerequisites are then assigned to an \code{MC} sampler,
which is used to generate probabilistic (i.e. including uncertaitainty quantification) forecasts.

Additionally, if a validation dataset is also available (i.e. not used for the preceeding inference),
it can be used for the evaluation of the error
in order to compute the predictive cross validation likelihood or distance.
For an example refer to \url{examples/randomwalk-forecast}.
An analogous SPUX configuration could also be 
obtained (see \autoref{s:propagation-forecasting})
by explicitly specifying prior distributions of parameters and intial state
instead of providing 'pastdir'.

\begin{figure}[!ht]
\centering
\includegraphics[width=\textwidth]{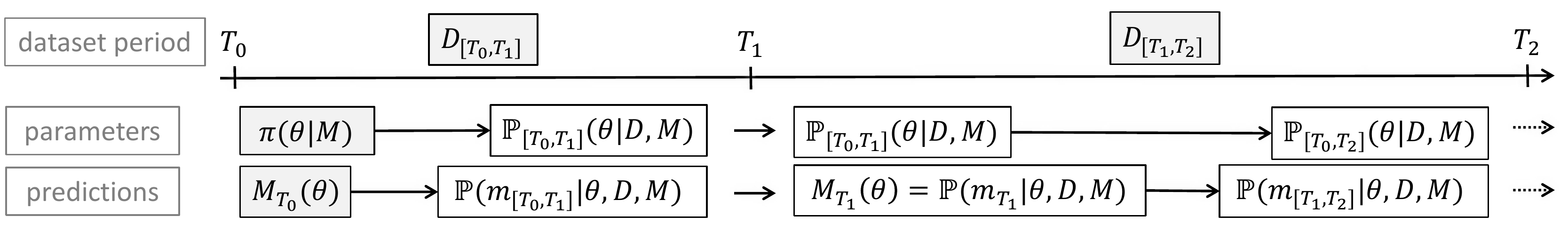}
\caption{Sequential Bayesian updating with SPUX for datasets split across multiple time periods.
Gray and white backgrounds indicate anticipated inputs and inferred outputs, respectively.}
\label{f:spux-framework-scheme-sequential}
\end{figure}

Secondly, upon aquisition of an additional dataset (for a time period $[T_{i}, T_{i+1}]$ following an already inferred time period $[T_{i-1}, T_{i}]$,
as depicted in \autoref{f:spux-framework-scheme-sequential}),
posterior distributions of model parameters and final model states at time $T_i$ 
can be used as prior distributions of model parameters and initial model state for the Bayesian inference within the succeeding time period $[T_{i}, T_{i+1}]$, respectively.
In such a case, the location of the "past" inference and the additional dataset (for period $[T_{i}, T_{i+1}]$) need to be specified;
unless explicitly specified otherwise, the error and units will be reused from the configuration of the specified "past" inference.
If model parameters are not expected to be influenced,
the model states for the next time period $[T_{i}, T_{i+1}]$ can be inferred by using the \code{PF} likelihood
while keeping the model parameters distribution unchanged by using the \code{MC} sampler
(see the example provided in \autoref{s:rw-forecast-assimilate-scripts} (\autoref{lst:spux-assimilate.cfg}) and \url{examples/randomwalk-assimilate}).
Alternatively, if not only the model state but also the model parameters need to be updated when processing the next time period $[T_{i}, T_{i+1}]$,
a full sequential Bayesian update can be configured analogously to the example provided in \autoref{s:usage-conf-file},
but with the \code{pastdir} specified instead of the \code{prior} and \code{initial}.
For an example, please refer to \url{examples/randomwalk-update}.

\section{SPUX framework usage and customization}
\label{s:customization}

This section starts with an overview of the available
framework setup options in \autoref{s:framework-setup}
including an overview of "sandboxing" strategies on \autoref{s:sandboxing},
and continues by describing key SPUX customization guidelines for the most common necessities. 
Those are: how to couple an application with the framework by defining a new model (\autoref{s:add:model}),
including potential options
to include stochastic processes for model input/parameters (\autoref{s:add:processes}); how
to handle model state serialization (\autoref{s:add:serialization}); how to
specify a prior distribution for all (model and observational error) parameters (\autoref{s:add:distribution});
how to define an error for the available dataset(s) (\autoref{s:add:error}).
In addition, \autoref{s:add:auxiliary} describes how optional "auxiliary" output and datasets,
that are not in a form of a \code{pandas.DataFrame},
can be incorporated into the model, error, and distribution classes.
Multiple independent datasets can be combined using the \code{Replicates} aggregator,
introduced in \autoref{s:replicates}.
Instructions on possible built-in parallelization techniques for Python applications
or, alternatively, on executing existing parallel user applications within the SPUX model environment
are presented in \autoref{s:parallel:model}.
Finally, \autoref{s:add:component} provides some guidelines on advanced customization options such as
writing a custom SPUX component (e.g. sampler, aggregator, likelihood, distance, etc.),
with an example of SABC sampler described in \autoref{s:rw-abc}.
%
SPUX documentation might incorporate
improvements made after the publication of this manuscript.


\subsection{Options for framework setup, component configuration and report}
\label{s:framework-setup}

Multiple (non-mandatory) options to configure
components, framework's "global" setup, and the report
can be specified directly in UI configuration file \code{spux.cfg} as indicated in
\autoref{lst:spux.cfg} and \autoref{lst:spux-options.cfg}.
In the following, we describe the corresponding API methods and list examples of such options.

In particular, an optional \code{configure(...)} method is available
for each component, implementing functionalities
usually shared by all the components of a given component type.
The following table provides a brief overview of all available \code{configure}-options (excluding
options already introduced in the preceeding sections) for model, likelihood/distance, and sampler component types:
\begin{center}
\begin{tabular}{r|c|l}
\hline
\code{templatedir} & \code{None} & directory with intial sandbox contents for the model\\
\code{statefiles} & \code{None} & sandbox files relevant to the model state (see \autoref{s:add:serialization-sandboxing})\\
\code{ignore} & \code{None} & list of non-serializable model attribute names (see \autoref{s:execution})\\
\code{timeset} & \code{4} & integer: points in-between dataset snapshots; iterable: predictions times\\
\code{auxset} & None & auxiliary observational datasets (see \autoref{s:add:auxiliary})\\
\code{lock} & \code{None} & batch index to lock sampler's feedback to likelihood or distance\\
\hline
\end{tabular}
\end{center}
An iterable (e.g.~list or array) of times $t_0, \dots, t_{\bar N}$ can be used for \code{timeset}
to select corresponding model outputs $y_{t_n}$ for later post-processing,
providing additional intermediate time points (among dataset snapshots)
for a larger temporal resolution (i.e.~with $\bar N > N$).

The framework itself can be customized
by the optional \code{spux.configure(...)} (see \autoref{lst:configure}) and the mandatory \code{spux.setup(...)} (see \autoref{lst:infer}) methods.
The following basic arguments (with default values and descriptions)
are available:
\begin{center}
\begin{tabular}{r|c|l}
\hline
\code{seed} & 0 & integer seed (for hierarchical seeding of RNG libraries)\\
\code{verbosity} & 2 & hierarchical verbosity level (integer) for SPUX components\\
\code{informative} & ["performance"] & to save: "performance" "timestamps" "infos" "rejections"\\
\code{sandboxdir} & \code{"sandbox"} & directory for the "root" sandbox (see \autoref{s:sandboxing})\\
\code{trace} & "none" & sandboxes to keep (if used): "none" "best" "posterior" "all"\\
\code{outputdir} & \code{"output"} & directory for SPUX output files (see \autoref{s:usage-setup-output})\\
\code{history} & "none" & store "statefiles"/"auxiliary": "none" "best" "posterior" "all"\\
\code{states} & False & store final model states for forecasting or sequential updating\\
\code{checkpoint} & 600 & minimal time period (in seconds) between checkpoints\\
\hline
\end{tabular}
\end{center}
Note, that including more keywords in \code{informative} will consequently also
increase expected inference runtime, required operational memory,
and the total size of written output files (see \autoref{s:usage-setup-output}).
To keep different "statefiles" copies and archived auxiliary model predictions,
a dictionary with
respective \code{"statefiles"} and \code{"auxiliary"} entries can be specified for \code{history}.
Refer to SPUX documentation for the advanced options
(\code{redirect}, \code{cache}, \code{setupdir})
and for the additional options of functions within the
\code{test.py}, \code{synthesize.py}, \code{report.py}, and \code{best.py} scripts.

\subsection{Sandbox - filesystems and post-execution accessiblity}
\label{s:sandboxing}

As indicated in \autoref{s:framework-setup},
the "root" sandbox directory \code{sandboxdir}
contains nested sandboxes
for each (if used) batch, chain, replicate, and model (particle or trajectory).
By default, the "root" \code{sandboxdir} is placed in (node-local) fast virtual node-local RAM-based Linux filesystem called \code{tmpfs}
(by default mounted at \code{/dev/shm}) to avoid network and I/O overheads.
If the amount of system memory is a limiting factor,
alternative (node-local) filesystems can be used to avoid network (but not I/O) overhead.
For inference runs on high performance computing clusters,
if neither option is possible,
(shared) "scratch" file system can be used instead
(with associated network and I/O overheads).
If sandboxes located on node-local (not shared) filesystems
are inaccessible and the functionality of the \code{history}
option is not sufficient,
the best (or any other) model trajectory
can be re-executed (even after inference)
within a specified accessible \code{trajectorydir}
for model's sandbox (see \autoref{s:rw-trajectory}).

\subsection{Adding a model}
\label{s:add:model}

In the most common use case scenario of SPUX,
a user will wrap an existing application as the SPUX \code{model}
either by configuring an existing SPUX \code{model}
or by implementing a new SPUX \code{model} as Python class.
To avoid confusion,
we will refer to a user's existing application (in any programming language) as the "application",
and to the (built-in or custom) Python class coupling such application to the SPUX framework as the "model".
In this section we
review available model testing routines and
discuss two scenarios in detail:
using the built-in \code{External} model to manage
the (appropriately modified) application
and writing a new SPUX model class to explicitly wrap an (unmodified) application.
In both cases, the incremental model execution must be possible,
i.e.~a corresponding output is required for each specified \code{time} from an increasing list of times.

\subsubsection{Model testing and dataset synthesis}
\label{s:add:model-testing}

Automatically generated \code{test.py} (and \code{synthesize.py}) scripts
(using the cleaned up SPUX UI \code{spux.cfg} file to only define the model component and its options)
could be used to continuously test the development of a new model class.
In \code{spux.cfg},
model parameters can be specified as the \code{parametersfile} option,
defining the path to a text file containing rows with parameter names and values (separated by some white space),
and an array of times (e.g.~using \code{utils.period(...)}) for model evaluation can be specified as the \code{snapshots} option.
The optional \code{synthesize.py} script
uses the specified (or drawn from the prior) exact model parameters
to generate exact model outputs
and selected observations (with \texttt{error}, if specified) at the specified snapshots,
including the corresponding \code{exact.py} and \code{dataset.py} scripts to load them
(for instance, to generate the configuration section of the SPUX report).
Such synthetically generated datasets
provide an invaluable resource for making sure the correctness of your implementation,
especially because the posteriors for model parameters (and outputs)
obtained from the Bayesian inference can be compared to their exact values.

\subsubsection{Using \code{External} model for user's application}
\label{s:add:external}

The \code{External} SPUX model relies on an application execution \code{command},
with which users are already familiar from using the console (shell),
making this a good starting option for the first coupling of the application to the SPUX framework.
In particular,
the application execution \code{command}
can be used to configure SPUX with an \code{External} model,
as indicated by examples in
\autoref{lst:spux-external.cfg} and \autoref{lst:configure-external.py}.

\noindent
\begin{minipage}[t]{.48\textwidth}
\begin{lstlisting}[caption={External model example in \code{spux.cfg}.},label={lst:spux-external.cfg},style=python]
model External
model.command r"./myapp <TIME>"
\end{lstlisting}
\vfill
\end{minipage}
\hfill
\begin{minipage}[t]{.48\textwidth}
\begin{lstlisting}[caption={External model example in \code{infer.py}.},label={lst:configure-external.py},style=python]
command = r"./myapp <TIME>"
model = External (command)
\end{lstlisting}
\end{minipage}
\\

%
The \code{External} model automatically
isolates the application to a unique sandbox directory,
where the corresponding initial model state is written to \code{initial.txt} file
if the \code{initial} model state is specified.
Subsequently, the \code{command} can be executed to evolve the current model state $m_{t_{n-1}}$ to the next $m_{t_n}$
by reading the automatically generated input files (\code{parameters.txt} for $\theta$, \code{time.txt} for $t_n$ and \code{seed.txt} for the seed),
and writing \code{output.txt} (see \autoref{s:outputs} for details).
Optionally, the contents of the corresponding files can be also passed to the application
as runtime arguments using \code{<PARAMETERS>}, \code{<TIME>} or \code{<SEED>} keywords (for substitution)
within the application \code{command}.
An alternative "direct" mode of the \code{External} model is available (see later sections for reasoning)
by setting model attribute \code{direct} to \code{True}. In this mode,
instead of executing the \code{command} sequentially for each required time $t_n$,
the \code{command} is executed only once to evolve model from the initial state $m_{t_0}$ to the final state $m_{T}$
by reading automatically generated input files \code{times.txt} and \code{seeds.txt}
(or using corresponding \code{<TIMES>} and \code{<SEEDS>} keywords),
and writing \code{outputs.txt}.

\subsubsection{Implementing a new SPUX model class - execution control}
\label{s:execution}

The functionality of the simple \code{External} model
might be insufficient; for instance,
it might be inadequate (binary executable),
inconvenient (requires changes in application interface)
or inefficient (requires storing model state inbetween \code{run(...)} calls)
to rely only on the application modifications.

\begin{figure}[!ht]
\centering
\includegraphics[width=\textwidth]{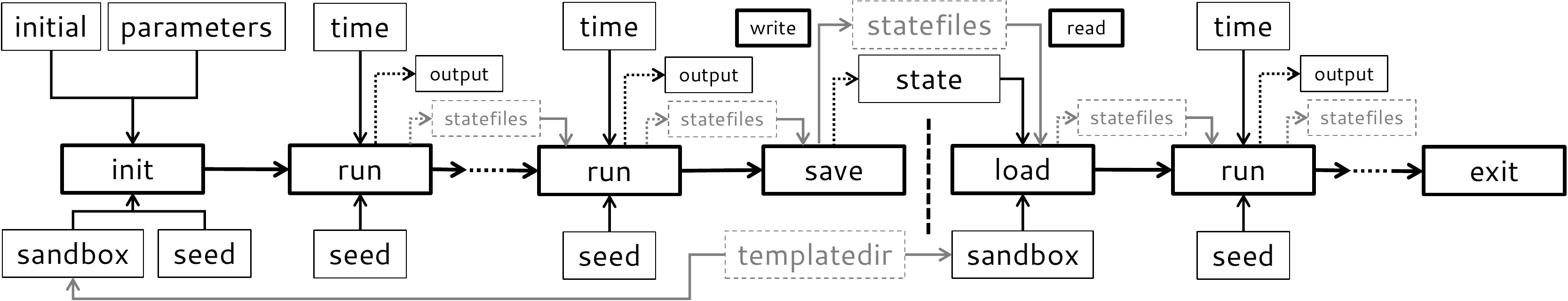}
\caption{Execution flow scheme for the SPUX model, controlling the user's application.
The middle row indicates order of model methods calls (with multiple \code{run} and \code{save/load} calls).
The top row indicates the arguments that are provided to these model methods,
and the bottom row indicates before which methods the built-in variables \code{sandbox} and \code{seed} are \emph{updated}.}
\label{f:spux-model-scheme-methods}
\end{figure}

A new SPUX model class can be written instead,
allowing unrestricted capabilities for interfacing
the application to the SPUX model,
possibly even without any modifications
to the application itself.
Such model class can be specified as the "model" component within
respective SPUX configuration files in \autoref{lst:spux.cfg} and \autoref{lst:configure}.
New model classes need to be derived from the base \texttt{Model} class
defined in the \code{spux.components.models.model} module.
%
The model execution flow scheme is provided in \autoref{f:spux-model-scheme-methods},
where arguments and built-in internal variables (introduced in \autoref{s:attributes})
are explicitly indicated for each model method.

A new SPUX model class needs to have an appropriately implemented \code{run(...)} method:
\begin{center}
\begin{tabular}{l|l}
\hline
\code{run(self,time)} & run model from current time ($t_{n-1}$) until \code{time} $t_n$, return model output $y_{t_n}$\\
\hline
\end{tabular}
\end{center}
In the method declaration above, \texttt{self} is a handle to the model class instance,
\texttt{time} corresponds to times in \code{snapshots}, \code{timeset}, \code{dataset} or \code{auxset},
and the model output (see \autoref{s:outputs}) consists of
a mandatory array of labelled values (and, if needed, an optional "auxiliary" object).
If \code{run(...)} fails (invalid parameters, invalid trajectory evolution, etc.),
nothing (\code{None}) can be returned instead of raising an error, if the user would like the inference to continue.
Optionally,
model initialization and finalization methods can be implemented:
\begin{center}
\begin{tabular}{l|l}
\hline
\code{init(self,initial,parameters)} & initialize model with \code{initial} $m_{t_0}$ and \code{parameters} $\theta$\\
\code{exit(self)}                    & finalize (cleanup) model after the last \code{run(...)} call\\
\hline
\end{tabular}
\end{center}
In some cases, performing the model evaluation $m \sim M(\theta)$
for all timesteps in a single function call (analogously to \code{direct} mode of the \code{External} model)
instead
of making incremental steps $m_{t_n} \sim \IP_n (\cdot|m_{t_{n-1}},\theta,M)$ for each time $t_n$
might be an easier way to wrap the user application and/or a faster way to execute the model
in cases where the incremental model execution is not required (currently it is required only by
the \code{PF} likelihood). This can be achieved by implementing a custom
\code{\_\_call\_\_(...)} method, returning model output for all \code{times} at once as a dataframe:
\begin{center}
\begin{tabular}{l|l}
\hline
\code{\_\_call\_\_(self,parameters,times)} & given \code{parameters} and \code{times}, return model output $y$\\
\hline
\end{tabular}
\end{center}
The built-in implementation of \code{\_\_call\_\_(...)} relies on using \code{init/run/exit} methods
and returns model outputs dataframe as well associated \code{info} and (optional) \code{timings} dictionaries (see \autoref{s:usage-setup-output}).
Additionally, \code{self.diagnostics} attribute might be set
(at runtime, by other components)
to a function returning a dictionary
of model output ($y_{t_n}$ at time $t_n$) diagnostics (e.g.~$\Pobs(y_{t_n}|\theta)$):
\begin{center}
\begin{tabular}{l|l}
\hline
\code{diagnostics(time,prediction,parameters,rng)} & return diagnostics of \code{prediction} $y_{t_n}$\\
\hline
\end{tabular}
\end{center}
The \code{info} in \code{\_\_call\_\_(...)}
then needs to be updated with returned diagnostics (if not \code{None}).

The implementation of model methods
could rely on direct imports of modules corresponding to existing applications written in Python,
could rely on customized versions of the corresponding methods found
in \code{External} or \code{Randomwalk} (or in the other models from the built-in examples),
or could be a combination thereof.
Three choices are available to control the application execution:
\begin{center}
\begin{tabular}{r|l}
\hline
basic & execute application command in a shell (an extension of the \code{External} model)\\
advanced & call application methods from Python using "drivers" (an efficient SPUX'onic way)\\
custom & implement any custom interface between the model and the application\\
\hline
\end{tabular}
\end{center}

The "basic" execution control
allows extending the capabilities of the
\code{External} model (see \autoref{lst:external} in \autoref{s:model-scripts}),
for instance, by storing model inputs in a different format,
reading model output from a custom output file,
supporting parallel applications (see \autoref{s:executors-model}),
implementing a custom \code{init(...)} method
to account for the specified initial model state.

The "advanced" and "custom" execution control allows the control of multiple execution stages in
the associated application directly from within the Python model methods,
avoiding the unnecessary overhead of application initialization and finalization
in-between consecutive calls of the model's \texttt{run(...)} method.
The computational efficiency gains are particularly large for a long time series datasets,
where \texttt{run(...)} needs to be called multiple times, and for models with small stochastic
volatility (including deterministic models), where model states filtering is infrequent (or absent)
inbetween consecutive \texttt{run(...)} calls.
In order to directly call routines within the user's non-Python application,
appropriate application-to-Python bindings can be used,
with built-in examples including:
\begin{center}
\begin{tabular}{l|l|r}
\hline
R & Python module "rpy2" & \url{examples/runoff}\\
Fortran & Python modules "f2py" or "ctypes" (C-ISO DLL) & \url{examples/superflex}\\
C/C++ & Swig (swig.org) code wrapper & \url{examples/hydro}\\
Java & Python module "JPype1" & \url{examples/IBM_2species}\\
\hline
\end{tabular}
\end{center}
In addition to these examples,
the most common practices of using the functionality of such Python bindings for SPUX
are combined into so-called "drivers" under \url{spux/drivers} directory.
Note, that loaded bindings or driver instances are usually non-serializable (see \autoref{s:add:serialization}),
and hence their names as model attributes (under \code{self}) need to be included in \code{ignore}, see \autoref{s:framework-setup}.
Next, we briefly describe built-in framework functionalities
that can be used in such a new model class.

\subsubsection{Model scope attributes}
\label{s:attributes}

Any model class instance has the following relevant internal attributes
accessible in all methods:
\begin{center}
\begin{tabular}{r|l}
\hline
\code{self.sandbox()} & path to an isolated sandbox directory (supports file name argument)\\
\code{self.parameters} & model parameters $\theta$ (available if \code{init(...)} is not specified)\\
\code{self.seed()} & list of integer seeds for each SPUX component (and iteration)\\
\code{self.seed.one()} & one integer seed, reduced from an array of hierarchical integer seeds\\
\code{self.rng} & \code{numpy.random.RandomState} as \code{random\_state} for \code{scipy.stats}\\
\code{self.verbosity} & integer indicating verbosity level, e.g. for custom \code{print(...)} usage\\
\code{self.print(string)} & print \code{string} to standard output, taking into account the verbosity\\
\code{self.replace(string)} & substitute \code{<PARAMETERS>}, \code{<SEED>} and \code{<TIME>} keywords in \code{string}\\
\code{self.shell(command)} & execute the specified \code{command} using shell in the sandbox directory\\
\hline
\end{tabular}
\end{center}
The sandboxing and seeding systems are described in more detail in the following sections.
Summary implementations of methods for an external application model (\code{External})
and a Python (\code{Randomwalk}) model
are provided in \autoref{s:model-scripts}, \autoref{lst:external} and \autoref{lst:randomwalk}, respectively.
For an alternative to \code{self.shell(command)} for parallel applications,
see \autoref{s:executors-model}.

\subsubsection{Model sandbox}
\label{s:add:sandboxing}

From within any method of the model, the path to the corresponding dedicated "local" sandbox can be retrieved by executing \code{self.sandbox()},
which might be different for \emph{each} \code{run(...)} call (e.g.~if \code{PF} likelihood is used).
This is a path to a unique working directory for each model class instance,
isolating multiple models executed in parallel
and preventing race conditions and conflicts for their input/output files.
Any file access within the model class can be conveniently performed
using the file path returned by \code{self.sandbox(filename)}.
If a user's model requires certain common files to be present in every "local" sandbox,
a template sandbox directory \code{templatedir}
can be specified in \code{model.configure(...)},
see \autoref{s:framework-setup} and \autoref{f:spux-model-scheme-methods}.
The \code{templatedir} is best placed
in parallel high performance "scratch" filesystems,
since contents of the template sandbox are automatically copied
(using efficient local caching) to each "local" sandbox.

\subsubsection{Model seeding}
\label{s:seeding}

Note, that \code{self.seed()}, \code{self.seed.one()} and \code{self.rng}
can be different for the initialization call \code{init(...)} and for \emph{each} \code{run(...)} call.
The \code{reseed()} method in the \code{test.py} script checks if the model supports such frequent updates of the seeding and of the random number generator.

\subsubsection{Initial model state (deterministic or stochastic)}
\label{s:initial}

Since the initial model state $M_{t_0}(\theta)$ can depend on the model
parameters, it is expected to be specified as a function \code{initial(parameters)},
the return value of which is passed to model's \code{init(...)} call.
Alternatively, \code{initial(parameters)} can return
a prior SPUX distribution for the initial model state $M_{t_0}(\theta)$,
a draw from which is passed to model's \code{init(...)} call.
In such case,
the posterior distribution of the initial model state
is then also inferred.

\subsubsection{Model output}
\label{s:outputs}

The \code{run(...)} method of the model needs to return labelled model output
$y_{t_n} = h (m_{t_n},\theta)$ constructed by calling the built-in \code{self.output(values,names,time)} method.
For the sake of simplicity and to keep the amount of data manageable,
only a list or an array of plain datatypes (e.g. floats, integers, strings, etc.) is allowed to be specified in the argument \code{values},
with corresponding names in a list (or an array) of strings specified in the argument \code{names}.
Alternatively, \code{values} can be a \code{pandas.DataFrame} with \code{labels} as index
and \code{values} as the first column.
For inference, all dataset (columns) labels
must be present among the model output labels.
For later reporting, the (extended) model ouput can contain
additional labelled quantities of interest from the full model state $m_{t_n}$.
Optional \code{\_\_call\_\_(...)} method of the model needs to return three objects:
a dataframe for \code{predictions} $y$,
an \code{info} dictionary with supporting information (including diagnostics, if requested),
and timing information (or \code{None}, if not available).
The \code{predictions} dataframe rows
need to be indexed by \code{times} and columns labeled by \code{names} -
for examples, see \code{Model} and \code{External} model classes.
For the (extended) model output of a stochastic model,
it could consist of multiple labelled quantities to enable, for example,
additional diagnostics plots to be included in the SPUX report (see \autoref{s:rw-report}):
\begin{center}
\begin{tabular}{r|l}
\hline
predictions2d & summary of statistical dependencies among (extended) model output $y_{t_n}$ pairs\\
prediction2d & joint posteriors for the most correlated (extended) model output $y_{t_n}$ pair\\
\hline
\end{tabular}
\end{center}

In some complex models (for instance, in computational fluid dynamics),
even the output prediction $y_{t}$ of the full state $m_{t}$
might consist of large multi-dimensional arrays (for instance, observed surface values)
instead of just a few scalar values.
In order to benefit from the built-in reporting capabilities
and at the same time have an efficient method for the evaluation of the error,
a diverse handful of important quantities of interest can be cherry-picked for the annotation
and the (remaining or entire) output prediction $y_{t}$
can be specified as "auxiliary" (see \autoref{s:add:auxiliary}).

\subsection{Processes (stochastic and deterministic)}
\label{s:add:processes}

Built-in processes for use within the SPUX model classes
are available in \code{spux.library.processes}.
For instance, \code{OrnsteinUhlenbeck} (i.e.~bounded standard Gaussian) process
with a temporal correlation length $\tau$ is available,
often used to facilitate inference of time-dependent input and/or parameters
for deterministic models \cite{reichert_2009_hydmodelunc}.
In such case, the model input and/or parameters
are stochastically
sampled from the process \code{evaluate(time,rng)} method
with \code{rng} set to model's \code{self.rng}.

\subsection{Model state serialization}
\label{s:add:serialization}

The "PF" likelihood estimator for stochastic models requires the user's model (and the underlying application) to have the capability
of cloning its state $m_s$ at any given time "snapshot" $s$ available in the specified dataset.
Additionally, the posterior forecast and sequential Bayesian updating described
in \autoref{s:propagation-forecasting} and \autoref{s:rw-forecast-sequential} also
rely on loading the final model state saved during the preceding inference run.
Such capabilities of the model are also tested by the automatically generated \code{test.py} script mentioned earlier.
In "clone" test,
the script runs the specified model up to the specified clone time
and makes a clone of the original model by saving its state.
Then, a second model is created by loading the saved state of the original model
and both models are run using the same seed up to the specified compare time.
The outputs of both models must be identical (up to numerical roundoff errors).
The "move" test checks if the model outputs
are consistent in case its sandbox is moved in-between consecutive model runs.
Note, that a freshly cloned model, resulting from the execution of the \code{load}
method as depicted in \autoref{f:spux-model-scheme-methods},
does \emph{not} execute the \code{init(...)} method;
in particular, any required corresponding application initialization procedures
need to be part of the \code{load} functionality.
Furthermore, application bindings or driver instances (see \autoref{s:execution}) are usually non-serializable
and hence their names (if assigned as models attributes)
need to be specified in \code{ignore} (see \autoref{s:framework-setup}).

In SPUX cloning is based on the concept of the model "state" serialization to a binary stream (array).
There are two potential sources for application's state information:
the model class instance and, if sandboxing is used, the contents of the associated sandbox directory.

\subsubsection{Model state serialization using "statefiles"}
\label{s:add:serialization-sandboxing}

If sandboxing is used and some files in sandboxes are relevant to the model state
(often corresponding to the "restart" or "pickup" files of the underlying application),
then the list of all such "statefiles" (wildcards such as "*" and directories are allowed)
must to be specified in \code{model.configure(...)}, see \autoref{s:framework-setup} and \autoref{f:spux-model-scheme-methods}.
Such "statefiles" might be dynamically generated during the \code{init(...)} and \code{run(...)}
methods of the model, and hence are not necessarily already present in the optional \code{templatedir}.
If the model state is completely determined by such sandbox "statefiles"
(i.e.~those files are all that is required for a successful call of the \code{run(...)} method, in addition to \code{templatedir} contents),
then the built-in model state serialization functionality is already sufficient,
independently of the application's programming language (including the \code{External} model from \autoref{s:add:external}).
However, writing relatively large "statefiles"
(in every \code{init(...)} and \code{run(...)})
as a strategy for model serialization might be inefficient
due to the resulting I/O and application initialization overheads
if the model state is needed only very infrequently
(or if \code{PF} is not used - for instance, during forecast runs).
To avoid such overhead,
instead of writing and reading "statefiles"
in \code{init(...)} and \code{run(...)} methods,
explicit model methods for "statefiles" writing and reading can be implemented by the user:
\begin{center}
\begin{tabular}{r|l}
\hline
\code{write(self)} & write "statefiles" representing model state $m_t$ to the model sandbox directory\\
\code{read(self)} & set model state $m_t$ by reading "statefiles" from the model sandbox directory\\
\hline
\end{tabular}
\end{center}
Note, that model sandbox is designed to contain only the "statefiles" corresponding to
the current model \code{time} $t_n$.
Any "statefiles" of the preceeding times (no longer required by \code{run(...)} method)
should be removed from the model sandbox.
If "statefiles" are actually required for all times (e.g.~in post-processing),
the \code{history} option needs to be set (see \autoref{s:framework-setup})
to enable automatic (within \code{self.output(...)}) "statefiles" copies to the output directory (see \autoref{s:usage-setup-output}).

\subsubsection{Model state serialization from model class instance}
\label{s:add:serialization-instance}

If, in addition to (or instead of) the sandbox "statefiles",
the model state depends on the information in the model class instance
(or in the memory of the driven application),
then the model serialization requirements depend on the application's programming language.
In particular, for (pickle'able) applications written in pure Python or R (using r2py bindings),
the built-in model state serialization functionality is already sufficient.
For applications written in other programming languages,
custom methods need to be implemented to
serialize the model into its binary representation (\code{state})
and to de-serialize such state into the model again
(including \code{write()} and \code{read()} for "statefiles", if used):
\begin{center}
\begin{tabular}{r|l}
\hline
\code{save(self)} & return a \code{bytearray} (binary array) as the current model state $m_t$\\
\code{load(self,state)} & load the specified model \code{state} $m_t$ represented in the \code{bytearray}\\
\hline
\end{tabular}
\end{center}
For some of the most common programming languages, built-in driver modules
described in \autoref{s:execution} can be used to implement the above model state saving and loading.

\subsection{Adding a distribution}
\label{s:add:distribution}

SPUX requires all joint statistical distributions,
such as model parameters prior $\pi(\cdot|M)$,
error $\Pobs(\cdot|M(\theta),\theta)$,
or model initial state prior $M_{t_0}(\theta)$
to be defined as a SPUX \code{Distribution}.
The mandatory functionality of a \code{Distribution} $X$
includes providing methods
\begin{center}
\begin{tabular}{r|l}
\hline
\code{(log)pdf(values)} & evaluate the joint (log) probability density $\IP(x)$ of values vector $x \sim X$\\
\hline
\end{tabular}
\end{center}
If the provided \code{values} are invalid,
0 and $-\infty$ should be returned, respectively.
Optional capabilities, such as marginal probability density
and quantile-quantile plots
or initial sampler parameters drawn from $\pi(\theta,M)$,
require implementation of additional \code{Distribution} methods:
\begin{center}
\begin{tabular}{r|l}
\hline
\code{(log)mpdf(label,value)} & evaluate marginal (log) probability density of specific $x$\code{[label]}\\
\code{intervals(alpha)} & support intervals (mass \code{alpha}) of marginal probability densities\\
\code{draw(rng)} & draw random values vector $x \sim X$ using the provided \code{rng}\\
\hline
\end{tabular}
\end{center}
Each element of the vector $x \sim X$ has an associated \code{label}, i.e.,
a string, with supported LaTeX syntax (for tables and figures),
such as, for example, \code{r'\${\textbackslash}mu\$'}, and an associated type
(by default, random variables in \code{Distribution} are of type \code{float}).
A dictionary of explicit types
for each random variable \code{label}
(for instance, loaded from a text file with rows of "name type" using \code{loader.read\_types(...)} from \url{spux.utils.io})
can be specified
in \code{configure(types=...)} of any distribution derived
from the base \code{Distribution} class. 
%
Highly complex distributions represented by hierarchical Bayesian networks,
where each random variable can have conditional dependencies on other variables via a directed acyclic graph,
can be also constructed as a SPUX \code{Distribution} by relying on already existing respective packages, such as \code{PyMC3} \cite{salvatier2016probabilistic}.
In the remaining of this section we describe how to construct a \code{Tensor} distribution from
univariate distributions of probabilistic libraries (such as \code{scipy.stats})
and outline available built-in transformation methods for further customization.

\subsubsection{Tensor distribution}
\label{s:add:distribution-tensor}

Already introduced in the \code{Randomwalk} model example in \autoref{s:usage-prerequisites},
a \code{Tensor} distribution class from \url{spux.library.distributions.tensor} module
provides the easiest and by far the most common way to specify a joint distribution
of statistically independent univariate random variables.
Only a \code{label}-indexed dictionary of required univariate distributions,
for example, constructed from univariate distributions of the \code{scipy.stats} module,
is needed to construct a \code{Tensor} distribution.

\subsubsection{Distribution transformations}
\label{s:add:distribution-transformations}

In some application scenarios,
either truncated distributions might be needed (for instance, for non-negative variables),
or some observations (the outputs and/or the dataset) need to be transformed
before the density of the distribution from the error can be evaluated
(for instance, in heteroscedastic errors).
For such purposes, the built-in transformation classes for univariate distributions
are available at \url{spux.library.distributions.utils}, with example usage in \url{examples/hydro/error.py}.
\begin{center}
\begin{tabular}{r|l}
\hline
\code{Truncate} & truncate tail(s) of a probability density function at the specified location(s)\\
\code{Concentrate} & concentrate tail(s) of probability density function to an atomic part\\
\code{Transform} & transform any continuous distribution by an invertible function\\
\hline
\end{tabular}
\end{center}

\subsection{Adding an observational error model}
\label{s:add:error}

In SPUX, an \code{error}
is defined using a function (or a callable object) which takes
model output prediction $y$ and parameters $\theta$ as arguments
and returns a corresponding SPUX distribution (see \autoref{s:add:distribution})
$D \sim \Pobs (\cdot|y,\theta)$.
Heteroscedastic errors
can often be easily implemented using the distribution transformation
described in \autoref{s:add:distribution-transformations}.
The SPUX framework aims at removing systematic bias in the
error by considering stochastic models instead of deterministic couterparts,
and hence by default temporally independent errors are assumed,
covering most of the realistic use cases.
If required, temporally correlated errors can be setup
by including past model outputs in \code{output}
and adding lagged dataframe columns in the dataset(s),
see \url{examples/superflex}.

\subsection{Adding auxiliary model outputs and observational datasets}
\label{s:add:auxiliary}

Arbitrary (\code{pandas.DataFrame}-incompatible) model outputs and datasets,
such as multi-dimensional arrays or unstructured relational sets,
can be also used in SPUX model, error, and distribution classes.
In particular, an arbitrary "auxiliary" object
can be passed in \code{run(...)} as part of the model output
using \code{output(...,auxiliary=...)}.
Correspondingly,
the \code{prediction.auxiliary} attribute will be available,
for instance, in the observational error model.
There, the SPUX distribution for "standard" observations
can be merged with a SPUX distribution (defined using \code{prediction.auxiliary})
for "auxiliary" observations
using the built-in \code{Merge} distribution from \code{spux.distributions.merge} module.
An \code{auxset} dictionary
with \code{"snapshots"} as time points
and a \code{"loader"} as function mapping a specified snapshot to the corresponding dataset object
(usually stored on a high-performance filesystem, see \autoref{s:launch})
needs to be specified, see \autoref{s:framework-setup}.
%
%
Note, that "auxiliary" model outputs are not returned by the model
\code{\_\_call\_\_(...)} method; if required internally, use the model \code{diagnostics} attribute functionality instead.
If \code{history} option is set (see \autoref{s:framework-setup}),
"auxiliary" model outputs
are serialized and packed into
archives in output directory (see \autoref{s:usage-setup-output}).

\subsection{Aggregating multiple datasets}
\label{s:replicates}

In some applications (for instance, \url{examples/hydro}),
multiple observational datasets are available for aggregation into the inference.
Each dataset corresponds to the same model parameters $\theta$,
but is a result of stochastic model evaluation with \emph{different} initial model states
and/or \emph{different} (mutually independent) stochastic trajectories.
The \code{Replicates} aggregator can be used to
aggregate datasets by packing the corresponding datasets in a dictionary indexed by dataset names.
The aggregator can optionally configure initial model states
from an analogously indexed dictionary
of corresponding \code{initial} functions,
errors
from a dictionary
of corresponding \code{error} functions,
and "auxsets"
from a dictionary
of corresponding \code{auxset} dictionaries.

\subsection{Parallel models and parallel user applications}
\label{s:parallel:model}

The example \code{Randomwalk} model introduced in \autoref{s:framework}
does not support parallelization, and hence no executor was attached to it.
However, an external user application might be very computationally expensive
and might be parallel (as a stand-alone executable or as a library)
or might be splittable into multiple independent tasks.
In this section we introduce built-in executors for already parallel user applications
and review different ways to support direct model parallelization in SPUX.
The supplementary testing and synthesis scripts mentioned earlier,
can also be launched for parallel models (i.e., with multiple workers attached),
as described in \autoref{s:launch}.

Since the application processes might be running on a different compute node,
in local filesystems the sandbox directory can only be used
for operations inside the SPUX model (not the coupled parallel user application).
If the sandbox directory is explicitly required inside parallel user application,
then the sandbox \emph{must} be located on a shared filesystem,
as described in \autoref{s:sandboxing}.

For applications parallelized using threads for shared memory architectures,
such as multi-core CPUs or GPUs,
it is sufficient (for any execution method from \autoref{s:execution})
to properly allocate enough computational resources
and properly bind SPUX MPI workers (ranks), such that
each application has a dedicated part of a multi-core CPU or a GPU.
If the (not yet parallelized) algorithms within model's
\code{init/run/\_\_call\_\_} methods
can be split into independent tasks,
consider distribution across parallel workers
of an attached "Pool" or "Ensemble" SPUX executor, see \autoref{s:parallelization}.

\begin{figure}[!h]
\centering
\includegraphics[width=\textwidth]{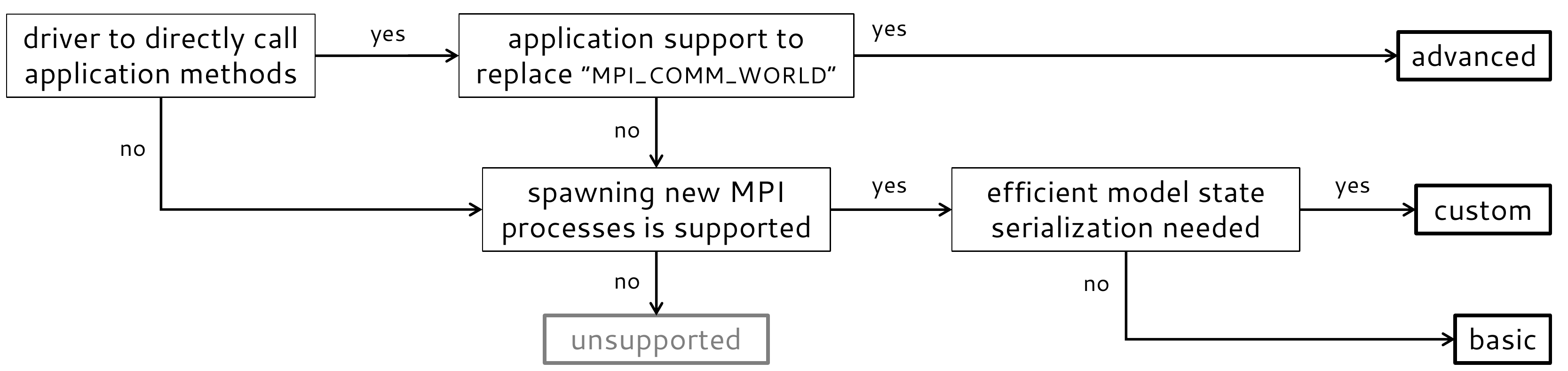}
\caption{A guideline scheme to select the appropriate parallel model executor mode
based on the characteristics of the application and of the MPI library.}
\label{f:spux-parallel-model-scheme}
\end{figure}

In the next sections we outline available SPUX functionalities
for applications parallelized using MPI (distributed memory parallelism),
where the built-in parallel model executor
is attached to the user's model with the specified number of workers.
In particular, "basic" mode replacing the built-in \code{self.shell(...)}
is described in \autoref{s:add:model-parallel-basic}.
Despite its generality and simplicity,
the "basic" model executor mode could be inefficient
due to multiple application execution calls.
Alternative model executor modes ("advanced" and "custom")
are introduced in \autoref{s:add:model-parallel-advanced}
and \autoref{s:add:model-parallel-custom} (with supplementary \autoref{s:add:model-parallel-app-comm}),
requiring only a single application launch (e.g.~loading a library or executing a process)
and relying on application execution control within \code{init/run/\_\_call\_\_} methods.
Such model executor modes allow one to avoid unnecessary filesystem related operations
and excessively large number of state saving/loading or "statefiles" writing/reading
by implementing custom \code{write/read} or \code{save/load} methods
(see \autoref{s:add:serialization}).
A guideline scheme for selecting the appropriate executor mode for the model
based on the available parallel features of the user's application and of the
MPI library is available in \autoref{f:spux-parallel-model-scheme}.

\subsubsection{Parallel model executor for parallel user applications - "basic" mode (execute)}
\label{s:add:model-parallel-basic}

As an extension of the "basic" execution mode in \autoref{s:execution},
the "basic" parallel model executor replaces
the built-in \code{self.shell(...)} in
\code{init/run/\_\_call\_\_} methods with
\begin{center}
\code{self.executor.execute (r'application', args = ['arg1', 'arg2'])}
\end{center}
This way, each application is executed (spawned) in parallel
(in the respective sandbox directory),
analogously to the manual launch using MPI.
For correct model serialization,
model (and application) state
within \code{init(...)} and \code{run(...)} methods
can be directly stored as model instance (\code{self}) attributes
and/or as files (e.g.~"statefiles") in its sandbox (without explicit \code{write/read} methods).
%

\subsubsection{Parallel model executor for parallel user applications - "advanced" mode (instruct)}
\label{s:add:model-parallel-advanced}

As an extension of the "advanced"
model execution method from \autoref{s:execution},
the "advanced" parallel model executor mode relies on direct
(but minimally intrusive) application control.
In particular,
on each parallel application worker in the \code{peers} intra-communicator,
a function to establish
and return (e.g.~using drivers, see \autoref{s:execution} and \autoref{s:model-executor-scripts})
an \code{interface} to the application
\begin{center}
\code{interface = method (manager, peers)}
\end{center}
can be specified for the built-in parallel model executor \code{establish(...)} method:
\begin{center}
\code{self.executor.establish (method)}
\end{center}
User-defined \code{instruction} functions
are required to call application method(s) on each parallel worker
(with the MPI intra-communicator \code{peers} instead of \code{MPI\_COMM\_WORLD})
using the binded \code{interface}:
\begin{center}
\code{result = instruction (interface, peers)}
\end{center}
SPUX model acts as a "manager"
to control application: specify parameters,
retrieve output,
and save/load model state or write/read "statefiles".
To achieve this,
custom \code{instruction} functions (see also 
\autoref{s:model-executor-scripts} for a practical example)
can be defined (even at runtime within model methods,
if specified model parameters, time, seed, etc.~need to be taken into account).
An \code{instruction} function can be dispatched from (any method of)
the SPUX model (as the "manager")
to all parallel application processes (as "workers") for execution
and retrieval of returned \code{instruction} results
by calling:
\begin{center}
\code{results = self.executor.instruct (instruction)}
\end{center}
Application \code{interface} can be terminated
by \code{self.executor.demolish()} in the model \code{exit()}.

\subsubsection{Parallel model executor for parallel user applications - "custom" mode (connect)}
\label{s:add:model-parallel-custom}

As an extension of the "custom" model
execution mode introduced in \autoref{s:execution},
a "custom" parallel model executor mode can be used
to launch (spawn) the application,
e.g.~in \code{init(...)}, with
\begin{center}
\code{self.executor.launch (r'application', args = ['arg1', 'arg2', '<PORT>'])}
\end{center}
where strings \code{<PORT>} among the list of arguments will be replaced
by the actual port, described in later \autoref{s:add:model-parallel-app-comm}.
Afterwards,
the manager-side (i.e.~SPUX model) MPI inter-communicator to the parallel workers (i.e.~user application)
can be obtained and released by, respectively:
\begin{center}
\code{workers = self.executor.connect()} and \code{self.executor.disconnect(workers)}
\end{center}
executor method calls in any model method, as many times as required.
Note, that the MPI communicators are not serializable and hence
must be added to the \code{ignore} list (see \autoref{s:framework-setup})
if stored as model (i.e.~\code{self}) attributes.
The manager-side MPI inter-communicator to workers
can be used for simulation control,
parameters specification, output retrieval,
saving/loading of the model state or writing/reading of "statefiles",
and for termination of application execution in model \code{exit()}.

\subsubsection{Application-side communicators for parallel user application in "custom" mode}
\label{s:add:model-parallel-app-comm}

In the "custom" model executor mode,
in addition to a standard intra-communicator among other application workers
(usually \code{MPI\_COMM\_WORLD}, as in a manual MPI run),
each parallel worker (of the user's application)
also retrieves a corresponding worker-side inter-communicator with the "manager" (the model)
by calling \code{MPI\_Comm\_Connect(port,...)}.
The \code{port} can be passed as the command line argument
(or, alternatively, as a file in the model sandbox) to the application executable.
The worker-side MPI inter-communicator to "manager" can be used 
to manage the execution flow within the user's application according to the requests from the
manager side (parameters acquisition, output reporting, saving/loading model state and writing/reading
"statefiles", etc.).

\subsection{Adding an arbitrary SPUX component}
\label{s:add:component}

In this section we briefly overview
guidelines for adding new SPUX component algorithms
(e.g. for sampler,
aggregator, likelihood or distance component types),
or even new SPUX component types.

A new SPUX component algorithm class
inherits the corresponding component type class,
and can potentially re-define mandatory
\code{require} dictionary and/or optional \code{configure} method (as in \code{configure.py}),
see \autoref{s:component-scripts}, \autoref{lst:component-algorithm}.
The \code{require} dictionary contains \code{"executor"}
with an attachable executor type name (from \autoref{s:executors-types} with default being "Pool")
and (optionally) \code{"tasks"} with a list of assignable component type names (absent by default).
Optional \code{init} and mandatory \url{__call__} methods of the component algorithm class
are responsible for the actual computations (as in \code{infer.py})
and can access \code{self.sandbox}, \code{self.seed} and \code{self.rng} instances.

A new SPUX component type class
(i.e.~a row in the components table, \autoref{s:specification})
inherits the \code{Component} class,
defines the mandatory \code{component} attribute (a string for type name),
and can re-define the \code{require} attribute
and the optional \code{configure} method,
see \autoref{s:component-scripts}, \autoref{lst:component-type}.
The \code{Component} class from the \code{spux.components.component} module
implements default attributes and methods common among all SPUX component type classes,
such as \code{name}, \code{require}, \code{evaluations},
\code{assign}, \code{attach}, \code{copy}, \code{setup}, \code{isolate}, \code{plant},
\code{spawn}, \code{consensus}, \code{feedback}, \code{feed}, \code{prepare}.

\subsection{Approximate Bayesian Computation type samplers}
\label{s:rw-abc}

\begin{figure}[!ht]
\centering
\includegraphics[width=\textwidth]{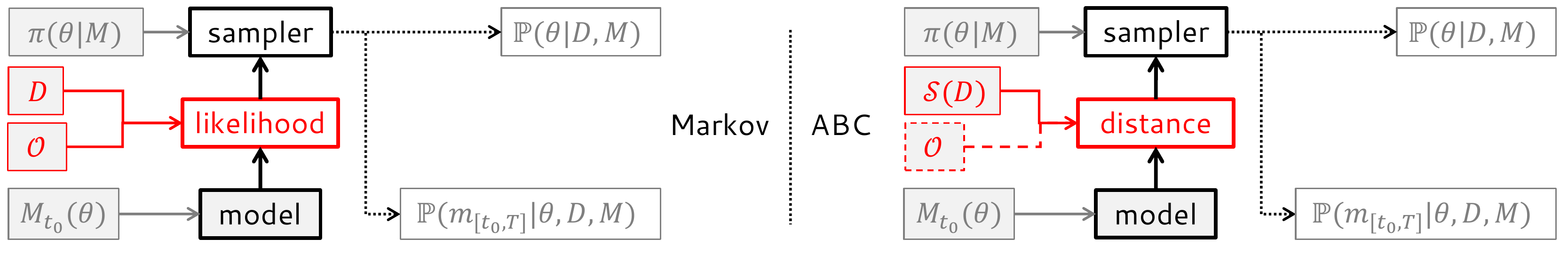}
\caption{Differences in the components configuration for the SPUX framework using ABC-type sampler (right part).
The components that are different from the Markov-type samplers (left part, refer to \autoref{f:spux-framework-scheme} for a detailed description)
are highlighted in red.}
\label{f:spux-framework-scheme-abc}
\end{figure}

In ABC-type samplers (introduced in \autoref{s:sampling-abc}), the (approximate-)likelihood based samplers are replaced
by approximate samplers based on an empirical distance between
the appropriate summary statistics $\mathcal{S}(y)$ of model output $y$ and 
$\mathcal{S}(D)$ of the observational dataset $D$,
as depicted in the adapted SPUX framework scheme in \autoref{f:spux-framework-scheme-abc}.
If, in addition, the (optional) observation error model is available,
then it will be incorporated into the model output before the distance evaluations.
Alternatively, the user's SPUX model might already contain all required uncertainties
arising from both the generative (deterministic or stochastic) model
and the stochastic error, in which case the observational errors
are also incorporated for the distance evaluations.
However, the explicitly defined error model (i.e., not part of the generative model)
allows for future forecast of true model output,
whereas hiding the error within the generative model
only allows for future forecast of observed (i.e., including observational noise) model output.
For the ABC-type samplers,
the number of required posterior samples
can be set by specifying the \code{batchsize}
for the sampler constructor,
and the number of iterations for convergence
can be set by specifying the \code{batches}
(instead of the \code{samples} for the Markov-type samplers)
for the call \code{sampler(...)}.
%
For an example using the SABC sampler, please refer to \url{examples/randomwalk-sabc}.

\section{SPUX framework parallelization}
\label{s:parallelization}

In this section we describe three built-in executor types
and their load balancing techniques.
For the design and implementation of the hierarchical parallelization sub-system
supporting multiple nested executors
for simultanous parallelization of multiple nested SPUX components,
see \autoref{s:executors-design}.

\subsection{SPUX executors - types}
\label{s:executors-types}

Any set of independent tasks within any of the SPUX components
can be executed in parallel using built-in SPUX executors.
The parallel design of SPUX is centered on three main types of
executors, each supporting different functionality
and (usually) meant to be used in different SPUX components:
\begin{center}
\begin{tabular}{r|l}
\hline
pool & dynamically executes a set of independent tasks; task "states" are discarded\\
ensemble & adaptively executes a set of independent ensemble tasks (maintaining task "states")\\
model & dynamically executes a user's application (discarding or maintaining its "state")\\
\hline
\end{tabular}
\end{center}
Executors can be attached to SPUX components without any restrictions on the executor type,
however, the list of available executor capabilities (methods) is checked
at the \code{attach(...)} method of the corresponding SPUX component.
As described in the tutorial, the default executors of each type are
the serial executors with rather straight-forward implementations,
namely \code{SerialPool}, \code{SerialEnsemble}, \code{SerialModel}.
Currently, parallel executors of each type are implemented in SPUX
using the Message Passing Interface (MPI) library \cite{MPI3}
via the object-oriented Python MPI bindings package \texttt{mpi4py} \cite{Dalcin2011}.
MPI dynamic process management for parallel workers is employed for deploying nested parallelization.
Pickle-based communication of arbitrary serializable Python objects
is used for execution workflow management across managers and workers,
as the memory and processing time overheads are negligible.
Additionally, for multiple subsequent executor calls (e.g.~for iterative sampling within sampler),
executor task caching allows to perform task serialization and communication only once.
Finally, memory-efficient raw binary arrays (\texttt{bytearray}) are used for efficient communication of model states.

\subsubsection{SPUX "pool" executors}
\label{s:executors-pool}

A "pool" type executor can be used by calling its \code{map(functions,parameters)} method,
where either argument (or both) can also be iterable (for instance, a list) over
corresponding parallel tasks.
%
For lower communication overhead, consider providing lists of \code{sandboxes} and/or \code{seeds}
as arguments in \code{map(...)} instead of using lists of \code{functions}.
An optional list of arguments can be specified in \code{map(...)} as \code{args}
and will be passed to the execution of the function(s) by \code{function(...,*args)}.

\subsubsection{SPUX "ensemble" executors}
\label{s:executors-ensemble}

An "ensemble" type executor does not accept tasks directly,
but requires an instance derived from the \code{Ensemble}
base class defined in the \url{spux.library.ensembles.ensemble} module.
Currently the only implemented ensemble class is for a \code{Particles} ensemble of SPUX models
(particles to be used in the \code{PF} likelihood).
For example, the "ensemble" executor, attached to an instance of the \code{PF} likelihood,
manages the parallel initialization, method execution and resampling of (indexed) ensemble tasks.
%
In particular, in-between the ensemble executor \code{connect(ensemble,indices)} and \code{disconnect()} methods,
\code{call(method,args)} and/or \code{resample(indices)} methods can be called multiple times,
each time executing the specified ensemble method and/or performing ensemble resampling, respectively.
In the \code{resample} method,
ensemble tasks can be cloned (duplicate indices) and deleted (missing indices),
including the load re-balancing across the resampled sub-ensembles.
A detailed description of the parallel resamplig procedure
is available in the earlier manuscript \cite{Sukys2017a}.

\subsubsection{SPUX "model" executors}
\label{s:executors-model}

A "model" type executor is needed to execute a parallel user application.
As described in \autoref{s:parallel:model},
three parallel model executor modes to manage parallel application workers are supported:
\begin{center}
\begin{tabular}{r|l}
\hline
basic & \code{execute(command)} by spawning a new process (analogously to \code{mpiexec})\\
advanced & \code{instruct(...)} to call \code{instruction(...)} with MPI intra-communicator \code{peers}\\
custom & \code{launch(application)} and \code{connect()/disconnect()} MPI inter-communicators\\
\hline
\end{tabular}
\end{center}
%
%
Options, such as \code{args} list of runtime arguments,
can be specified in \code{execute(...)} and \code{launch(...)} methods.
%
%
For serial executors (\code{workers} set to \code{None} or \code{0}),
both methods fall back to \code{shell(...)}.

\subsection{SPUX executors - load balancing}
\label{s:executors-balancing}

In all parallel executors, a collection of multiple tasks must be carefully
dispatched to (or re-distributed among) parallel workers to ensure
the minimal needed wall-clock runtime to execute all tasks.
In this section we describe such load balancing algorithms for the pool and ensemble type executors.
For "model" type executors, the load balancing is federated to the user's application itself.

\subsubsection{Load balancing for pool-type executors}
\label{s:executors-balancing-pool}

All tasks requested in the "map" method of the "pool" type executor are executed dynamically,
with pending tasks being constantly dispatched to workers as they become available.
For parallel runs with multiple datasets (e.g. \url{examples/hydro} example),
the \code{Replicates} aggregator performs guided load balancing
by evaluating the lengths of the associated datasets and sorting component evaluations,
taking into account the component adaptivity as well
(for instance, the replicate-dependent adaptive number of particles in the \code{PF} likelihood).
Higher priorities are assigned to components with longer datasets
and large number of particles (if applicable),
and lower priorities are assigned to the components with shorter datasets
and smaller number of particles (if applicable).

\subsubsection{Load balancing for ensemble-type executors}
\label{s:executors-balancing-ensemble}

During routing, re-sampled tasks are instructed to be moved
from over-utilized workers to the neighboring under-utilized workers across their intra-communicator.
Routing objects include worker-specific information regarding task
identification, source (present worker address), destination (re-routed worker address), re-identification (for post-routing re-seeding), and affinity (local or remote node).
An empirical greedy algorithm is employed for constructing routings based on
re-sampled task counts, current task distribution across workers, and the maximal worker load.
For each worker, resident re-sampled tasks are kept in place
provided the worker loads do not exceed the maximal worker load.
Remaining tasks are routed either to a under-utilized worker on the same compute node (determined by affinity),
or, in the worst case, to the closest (w.r.t. proximity of worker ranks) under-utilized remote worker.
To further reduce communication overhead, identical tasks are moved together
by moving only a single particle and then replicating it on the destination worker.
Such load balancing strategy, however, does not take into account possibly heterogeneous model runtimes for
each task, and hence there could be potential gains from dynamic balancing frameworks using task stealing
instead of an adaptive routing approach.
Sandboxes, associated to the corresponding tasks,
are assumed to be isolated from each other, unless local affinity (determined automatically) is set in the routing information.
For local affinity, sandboxes exist on the same compute node (with access to the same filesystem), and hence a
simple and efficient sandbox renaming (internally consisting of stashing and fetching) is sufficient.
Such "hybrid" parallelization, inspired by existing MPI + OpenMP paradigms,
harnesses the node-local connectivity to minimize the required sandbox-related communication.
For remote affinity, sandboxes are "pseudo-moved" from one remote filesystem to another,
by copying template sandbox on the destination, and, if sandbox "statefiles" are specified,
saving the sandbox state, sending it to the destination, and loading it into the newly created sandbox.
After removal of all extinct particles,
at the expense of memory usage the resampling process prioritizes
overlapping the communication synchronization overhead
with any other pending task that does not require information exchange
with other parallel ensemble workers.
In particular,
after all asynchronous sends and receives for task exchange according to routings are launched,
all orphan particles (already sent, but not to be kept) are removed,
remaining particles are stashed,
ensembles are synchronzed to prevent race conditions,
stashed particles are fetched using re-identification,
local particles are replicated,
and only in the last step received remote particles are stored and replicated,
while still waiting in the background for all local particles (if any) to be sent.
For a more detailed description and an illustrative scheme
of such adaptive ensemble task resampling,
refer to the earlier manuscript \cite{Sukys2017a}.

\section{Future developments}
\label{s:summary}


In the near future,
multi-level uncertainty quantification methods (e.g. ML(ET)PF, MLCV) will be investigated.
These methods can greatly reduce the amount of required computational resources
for Bayesian inference and forecasting by configuring applications
for multiple different accuracies (e.g. resolutions).
These capabilities will be offered in SPUX through the the foreseen \code{MLCV}
aggregator component.
Furthermore, samplers optimized for multi-modal posteriors will be investigated.
%
%
Additionally, shared memory optimizations for nodel-level model state communication,
dynamic balancing using task stealing in ensemble tasks resampling,
and machine learning based dynamical process scheduling for pool executors
are foreseen as further improvements of an already very well performing SPUX parallelization
suite. Depending on the level of support for general purpose distributed tasking libraries,
new parallel executors could be added.
%

\section{Author contributions and acknowledgments}

\textbf{Author contributions.}
J\v{S} designed and implemented the prototype framework, plotting,
automatic report generation, compiled the documentation,
contributed examples for built-in and "hydro" models, and led the preparation of this manuscript.
J\v{S} and MB co-designed and implemented the parallelization specification, all main SPUX components,
and SPUX interfaces (UI and API).
MB also designed and implemented the legacy MPI connector,
distribution variable types,
and contributed examples for "superflex" and "ibm" models.
\\
\\
\noindent
\textbf{Acknowledgments.}
Authors would like to thank
Uwe Schmitt and Mikołaj Rybiński (ETH Zurich) for support
with CI/CD and unit testing on EULER and Daint clusters at the Swiss Supercomputing Center (CSCS),
Artur Safin, Marco Dal Molin and Lorenz Amman (Eawag) for contributing example models,
Mira Kattwinkel (U Koblenz-Landau) for helping to design the initial SPUX framework prototype,
Alexander Madsen (PSI), Anthony Ebert (USI and Eawag), Simone Ulzega (ZHAW) for contributing and testing SABC sampler,
Peter Reichert (Eawag), Andreas Scheidegger (Eawag), Carlo Albert (Eawag), Fabrizio Fenicia (Eawag), Vilma {\v S}ukien{\. e}
for providing feedback to this manuscript and framework usage,
Panagiotis Hadjidoukas (IBM Research Zurich) for advice regarding task-based parallelism design,
and Siddhartha Mishra (ETH Zurich) for access to EULER cluster.
\\
\\
\noindent
\textbf{Funding.}
We also acknowledge the Eawag Directorate Discretionary Funding for granting financial resources
and Swiss Supercomputing Center (CSCS) for granting computational resources in the development project d97.

\bibliographystyle{alpha}
\bibliography{bibliography/library,bibliography/library-jorg,bibliography/library-nele,bibliography/library-mira,bibliography/library-peter}

\appendix

\section{SPUX setup and output}
\label{s:usage-setup-output}

The \url{setup} directory
contains generated raw (binary, text, and LaTeX) information regarding SPUX framework configuration and setup,
corresponding to \code{configure.py} and \code{infer.py}, respectively.
%
%
The \url{output} directory is generated incrementally using checkpointing and
contains multiple subdirectories:
\begin{center}
\begin{tabular}{r|l}
\hline
\code{samples} & samples of posterior parameters, outputs, supporting "infos", timings\\
\code{best} & the "best" trajectory (parameters and output); also see \autoref{R-t:randomwalk_best-specification}\\
\code{diagnostics} & diagnostics information (e.g.~metrics)\\
\code{specifications} & structure specifications of the outputs/infos files and the "best" trajectory\\
\code{pickup} & framework (e.g.~sampler) pickup information (for \code{{-}{-}continue})\\
\code{states} & final states samples of posterior model trajectories\\
\code{statefiles} & (only if used) copies of "statefiles" for all times \\
\code{auxiliary} & (only if used) auxiliary model outputs\\
\hline
\end{tabular}
\end{center}
Contents of the \code{samples} subdirectory can be loaded and used for any custom post-processing:
\begin{center}
\begin{tabular}{r|l}
\hline
\url{parameters-*.csv} & a CSV file containing posterior samples of model parameters\\
\url{predictions-*.dat} & a binary file containing (posterior) of model output samples\\
\url{infos-*.dat} & a binary file containing supporting information (e.g. acceptance rates)\\
\url{timings-*.dat} & a binary file containing framework timings (runtimes and timestamps)\\
\hline
\end{tabular}
\end{center}
Refer to \code{report.py} for an example of loading all samples files into dataframes (for CSV)
and lists of dictionaries using \code{loader.reconstruct(...)} from \code{spux.utils.io}.
The corresponding specifications tables
are also listed in the \autoref{R-s:report-specifications} of the SPUX report
as \autoref{R-t:randomwalk_predictions-specification}, \autoref{R-t:randomwalk_infos-specification},
and \autoref{R-t:randomwalk_timings-specification}.
In particular, each model predictions (output) sample is a dictionary of either dataframes (default)
or \code{Trajectories} from \code{spux.utils.trajectories} module
(using memory-efficient compressed data structures \cite{jacob2015}, e.g.~for \code{PF}).

Infos contain additional (component-specific) information
for "infos" keyword requested in \code{informative}
and/or for a manually specified component
(such as \code{PF} likelihood or \code{Norm} distance) \code{diagnostics} list.
Unless "rejections" keyword is requested in \code{informative},
model outputs and infos corresponding to the rejected samples
are excluded (set to \code{None}),
and the estimates of prior and likelihood/distance are overwritten by the values of the accepted samples.

%
To ensure consistency, we strongly advice to access any SPUX framework output
(or configuration and setup options) using the SPUX \code{status},
which can be constructed using the \code{Status} class from the \code{spux.utils.status} module,
as is done in \code{report.py}.
Attribute \code{status.parameters} is a \code{pandas.DataFrame} loaded directly from \url{parameters-*.csv},
and additional \code{status} methods can be called with any required \code{batch}/\code{chain}/\code{time} arguments:
\begin{center}
\begin{tabular}{r|l}
\hline
\url{info(key,...)} & respective \code{key} value from loaded \url{infos-*.dat}\\
\url{prediction(...)} & respective model prediction from loaded \url{predictions-*.dat}\\
\url{timing(...)} & respective timing measurement from loaded \url{timings-*.dat}\\
\url{sandbox(...)} & respective model sandbox path (relative to the "root" sandbox)\\
\url{auxiliary(...)} & respective "auxiliary" model output (see \autoref{s:add:auxiliary})\\
\url{statefiles(...)} & respective paths to directories storing all "statefiles" (see \autoref{s:add:serialization-sandboxing})\\
\hline
\end{tabular}
\end{center}
Internally, these methods use \code{"origins"} stored in sampler "info" to locate the corresponding accepted samples
and functionality of the \code{Trajectories} class to follow trajectory filtering (if PF was used).
%

The \url{figures} directory contains all raw figure files
in multiple formats (default: PDF (vector), PNG (raster))
and with the associatiated caption files \url{*.cap}
containing the description of the figure contents.
The contents of the \code{setup} and \url{figures} directories are used
to generate the LaTeX source files in the \url{report} directory,
which are compiled into the SPUX report (A4 and "slides" layouts).

\section{Remark on workers for parallel SPUX executors}
\label{s:remark-workers}

The freedom to attach arbitrary many parallel workers for every SPUX component provides a lot of flexibility,
but also leaves ample space for computationally inefficient parallel configurations.
Hence, for large production runs, we strongly advice to follow these two guidelines:
\begin{itemize}
\item Attach most workers to the outer-most SPUX component(s) (e.g. \code{sampler}).
\item Avoid few parallel workers (less than 4) - replace them with the default \code{workers = None}.
\end{itemize}
SPUX will report the percentage of the number of manager cores w.r.t. the total number of cores.
%
Additionally, parallel performance and scaling plots
could be used to investigate
the model runtimes homogeneity and the synchronization overheads.
For highly heterogeneous model runtimes
(for instance, when model runtime strongly depends on the proposed model parameters),
or for the cases where sampler often proposes parameters outside the specified prior parameters
distribution, consider attaching more workers to the \code{PF} likelihood instead in order to
avoid very few tasks for each parallel worker in sampler (and aggregator) SPUX components.

\section{Example forecast and sequential assimilation scripts}
\label{s:rw-forecast-assimilate-scripts}

\noindent
\begin{minipage}[t]{.47\textwidth}
\begin{lstlisting}[caption={Example \url{spux.cfg} for forecast},label={lst:spux-forecast.cfg},style=python]
model Randomwalk
model.dt 0.1
aggregator Trajectories
aggregator.trajectories 64
sampler MC
sampler.chains 64
sampler.samples 500
pastdir "../randomwalk"
timeset utils.period (25, 30)
\end{lstlisting}
\vfill
\end{minipage}
\hfill
\begin{minipage}[t]{.47\textwidth}
\begin{lstlisting}[caption={Example \url{spux.cfg} for assimilation},label={lst:spux-assimilate.cfg},style=python]
model Randomwalk
model.dt 0.1
likelihood PF
likelihood.particles 256
sampler MC
sampler.chains 64
sampler.samples 1000
pastdir "../randomwalk"
dataset dataset-T_1-T_2.py
\end{lstlisting}
\end{minipage}

\section{Example model scripts}
\label{s:model-scripts}

\noindent
\begin{minipage}[t]{.49\textwidth}
\begin{lstlisting}[caption={External model methods},label={lst:external},style=pythonsmall]
from spux.utils.io import loader
from spux.utils.io import dumper
def run (self,time):
  p = self.sandbox("parameters.txt")
  t = self.sandbox("time.txt")
  s = self.sandbox("seed.txt")
  dumper.txt(p,self.parameters)
  dumper.txt(t,self.time)
  dumper.txt(s,self.seed())
  cmd = self.replace(self.command)
  code = self.shell(cmd)
  if code is None: return None
  o = self.sandbox("output.txt")
  return self.output(loader.txt(o))
\end{lstlisting}
\vfill
\end{minipage}
\hfill
\begin{minipage}[t]{.49\textwidth}
\begin{lstlisting}[caption={Randomwalk model methods},label={lst:randomwalk},style=pythonsmall]
def init (self,initial,parameters):
  self.d = parameters["drift"]
  self.v = parameters["volatility"]
  self.m = initial["position"]
  self.time = initial["time"]

def run (self,time):
  while self.time < time:
    dt = min(time-self.time,self.dt)
    n = self.rng.normal()
    self.m += dt*self.d
    self.m += sqrt(dt)*self.v*n
    self.time += dt
  return self.output([self.m],['p'])
\end{lstlisting}
\end{minipage}



\section{Example parallel model executor scripts for Fortran}
\label{s:model-executor-scripts}

In \autoref{lst:model-executor-establish},
the \code{application} module is obtained by compiling user's Fortran application with \code{f2py}.
On parallel application workers, the returned \code{interface} is passed to \code{instruction} in \autoref{lst:model-executor-instruction}.

\noindent
\begin{minipage}[t]{.47\textwidth}
\begin{lstlisting}[caption={Application interface example},label={lst:model-executor-establish},style=pythonsmall]
def interface (manager,peers):
  import application
  interface = application
  return interface
\end{lstlisting}
\vfill
\end{minipage}
\hfill
\begin{minipage}[t]{.47\textwidth}
\begin{lstlisting}[caption={Executor instruction example},label={lst:model-executor-instruction},style=pythonsmall]
def instruction (interface,peers):
  comm = peers.py2f()
  result = interface.main(comm)
  return result
\end{lstlisting}
\end{minipage}

\section{Example SPUX component scripts}
\label{s:component-scripts}

\noindent
\begin{minipage}[t]{.47\textwidth}
\begin{lstlisting}[caption={Component algorithm class example},label={lst:component-algorithm},style=pythonsmall]
class NewAlgorithm (NewType):
  def init (self,...):
    ...
  def __call__ (self,...):
    ...
\end{lstlisting}
\vfill
\end{minipage}
\hfill
\begin{minipage}[t]{.47\textwidth}
\begin{lstlisting}[caption={Component type class example},label={lst:component-type},style=pythonsmall]
class NewType (Component):
  component = "new_type"
  requires = {"executor":"Pool"}
  def configure (self,...):
    ...
\end{lstlisting}
\end{minipage}

\section{Adaptivity in the Particle Filter}
\label{s:pf-adaptivity}

For the SPUX framework, we are also introducing an empirical
adaptivity technique to dynamically control the number of particles used for the PF likelihood.
The proposed technique, as already common among complex adaptive methodologies such as \cite{elvira},
relies on empirical metrics; in this case, on
"fitscore", which describes the convergence progress of the posterior sampling,
and on "accuracy", which provides an estimate for the statistical error in the estimated likelihood.

\subsection{Fitscores of model parameters}
\label{s:pf-adaptivity-fitscores}

A fitscore indicator is simply the likelihood
normalized (purely for easier interpretation and generalization)
w.r.t. the dataset length,
dimensionality of the observations,
and the maximum of the corresponding error distribution densities.
As such, fitscore measures (analogously to likelihood)
how consistent the model output is
compared to the observational dataset.
Recalling the notations
\begin{equation}
\label{eq:O-notations}
\Pobs_{n}(D | y,\bth) = \Pobs(D_{t_n} | y_{t_n},\bth),
\quad
\Pobs_{n}^p(D | y,\bth) = \Pobs_{n}(D | y^p,\bth),
\quad
\Pobs_{n}(D | y,\bth) = \frac{1}{P}\sum_{p=1}^P \Pobs_{n}^p(D | y,\bth),
\end{equation}
we define fitscore as the average (over multiple dataset points $N$ and multiple particles $P$ of the Particle Filter)
logarithm of the normalized (with respect to probability density function value of the model prediction and the dimensions $d$ of the observations)
residuals (observational model error probability density of posterior model output):
\begin{equation}
\label{eq:fitscores}
r(\bth, D, M) = \frac{1}{N}\sum_{n=1}^N \frac{1}{P}\sum_{p=1}^P
\frac{\log\Pobs_{n}^p(D | y,\bth) - \log\Pobs_{n}^p(y |y,\bth)}{d}.
\end{equation}
The numerical value of the fitscore empirically tracks progress of the sampler convergence,
with higher values indicating that the sampler is most probably out of burn-in phase.

\subsection{Accuracies of likelihood estimates}
\label{s:pf-adaptivity-accuracies}

The variance $\sigma^2 (\log \hat L)$ of the estimated marginal log-likelihood $\log\hat L$
was found to be an important indicator controlling the convergence of the Markov chain sampler \cite{Kattwinkel2017}.
It can be bounded (with equality only for statistically independent $\Pobs_n$ over all $n = 1, \dots, N$) by the sum of log-likelihood variances from individual snapshots
(for brevity, we drop $(D |y,\bth)$ notation from $\Pobs_n^p$ and $\Pobs_n$):
\begin{equation}
\label{eq:var-logL}
\sigma^2 (\log \hat L)
=
\sigma^2 \left(\log \left( \prod_{n=1}^N \Pobs_{n} \right)\right)
=
\sigma^2 \left(\sum_{n=1}^N \log \Pobs_{n}^p\right)
\leq
\sum_{n=1}^N \sigma^2 (\log \Pobs_{n}).
\end{equation}
Motivated by this upper bound,
the empirical "accuracy" of the PF likelihood is defined as
the average (over multiple dataset points) standard deviations
for the estimated observational log-errors
(treated as random statistical estimates depending on $p$):
\begin{equation}
\label{eq:accuracies-implicit}
\delta(\bth, D, M) = \frac{1}{N}\sum_{n=1}^N \sigma (\log\Pobs_n).
\end{equation}
The numerical value of the accuracy empirically measures the
statistical accuracy of the marginal likelihood estimator computed by the Particle Filter,
normalized with respect to the length $n$ of the dataset.
The standard deviations $\sigma (\log\Pobs_n)$
for the estimated observational log-errors of the current snapshot,
are dynamically (i.e., during runtime) estimated
using 1st order Taylor-series approximation and the central limit theorem.
In particular, the variance of the logarithm of a random variable $a$
can be approximated using Taylor series around the mean $\mu(a)$:
\begin{equation}
\label{eq:var-log}
\sigma^2 (\log a)
\approx
\frac{\sigma^2 (a)}{\mu^2(a)},
\quad
\textnormal{provided}
\quad
\sigma^2 (a) \ll \mu^2(a).
\end{equation}
The summands $\sigma (\log\Pobs_n)$ in equation \ref{eq:accuracies-implicit}
can be approximated by such Taylor-series by setting $a = \Pobs_n$
and then applying a central limit theorem to approximate $\mu(\Pobs_n)$ and $\sigma(\Pobs_n)$
by $\hat\mu(\Pobs_n)$ and $\hat\sigma(\Pobs_n)$:
\begin{equation}
\label{eq:a}
\mu (\Pobs_{n}) = \mu (\Pobs_{n}^p) \approx \hat\mu(\Pobs_{n}^p)
=
\frac{1}{P} \sum_{p=1}^P \Pobs_{n}^p = \Pobs_{n},
\quad
\sigma^2 (\Pobs_{n})
=
\frac{\sigma^2 (\Pobs_{n}^p)}{P}
\approx
\frac{\hat\sigma^2 (\Pobs_{n}^p)}{P}.
\end{equation}
In \autoref{eq:a}, $\hat\mu(\Pobs_{n}^p)$ and $\hat\sigma(\Pobs_{n}^p)$ are the empirical estimators
for the mean and the variance as computed from the available $p$ samples $\Pobs_{n}^1, \dots, \Pobs_{n}^P$
of the random variables $\Pobs_{n}$.
Applying both approximations, the final estimate for the "accuracy" is then given by
\begin{equation}
\label{eq:accuracies}
\delta
\approx
\frac{1}{N}\sum_{n=1}^N \frac{\sigma (\Pobs_n)}{\mu(\Pobs_n)}
\approx
\frac{1}{N}\sum_{n=1}^N \frac{1}{\sqrt{P}} \frac{\hat\sigma (\Pobs_{n}^p)}{\Pobs_n}.
\end{equation}

\subsection{Adaptivity procedure and customization}
\label{s:pf-adaptivity-procedure}

Initially,
all particle filters
start with the minimal number of particles
(default is 4 or the number of parallel workers for \code{PF}).
Fitscore values above the customizable threshold (with the default value at -2)
indicate that the sampler is most probably out of the initialization phase,
and hence the particle filter adaptivity is activated.
The number of particles is then halved or doubled in each likelihood estimation step,
depending on whether the accuracy is above or below the requested accuracy envelope.
Requested accuracy envelope is determined by the accuracy and margins
specified within the adaptive PF likelihood.
The maximum number of particles is given by \code{particles} option of \code{PF}.
To guarantee the convergence of the posterior,
the particle adaptivity is suspended
after the user-specified "lock" sample batches.


\section{SPUX executors - design}
\label{s:executors-design}

\begin{figure}[!ht]
\centering
\includegraphics[width=\textwidth]{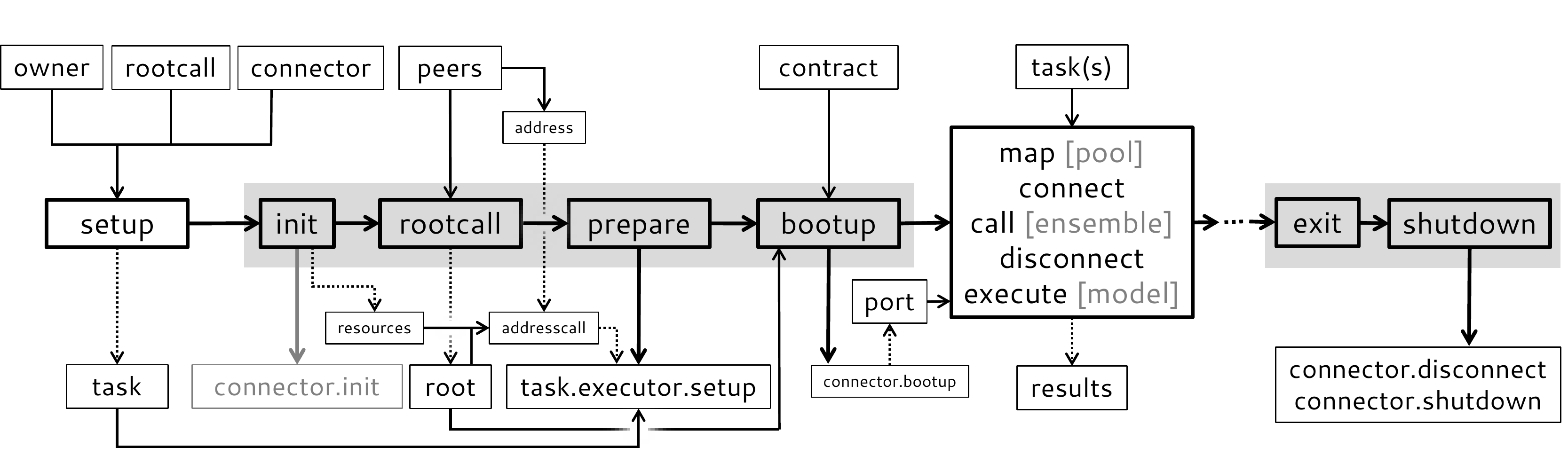}
\caption{Execution flow scheme for the SPUX executor base class (the manager side).
The thick solid arrows represent the sequence of the called methods
(the gray arrow indicates a call at the top-level executor only),
the thin solid arrows represent the required inputs,
and the dotted arrows represent the outputs (or inputs provided to subsequently called methods).
The tasks-to-results pipeline can then be executed multiple times,
with the final executor \code{exit()} deallocating all computational resources.}
\label{f:spux-executor-scheme}
\end{figure}

\begin{figure}[!ht]
\centering
\includegraphics[width=\textwidth]{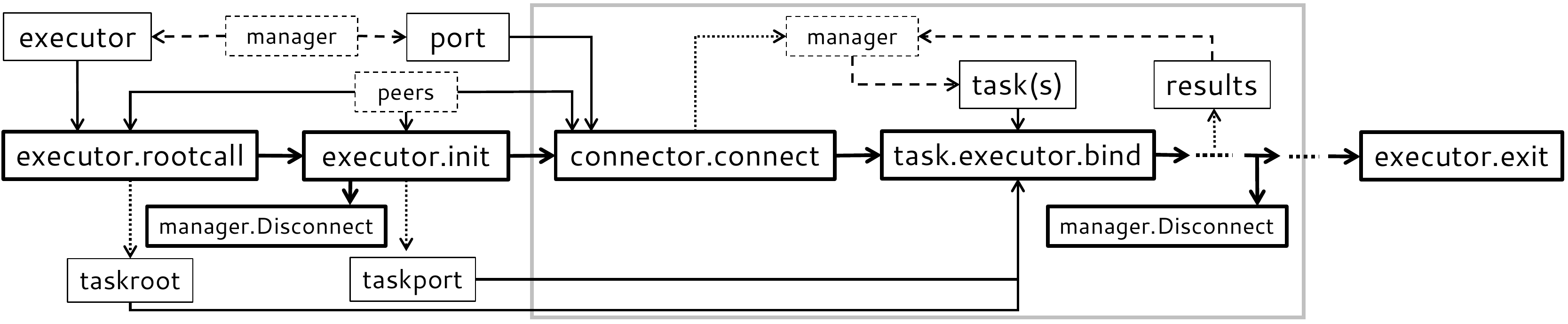}
\caption{Execution flow scheme for the SPUX contract (the worker side) required in SPUX executors.
The boxes with dashed outline represent the communicators and the dashed arrows
represent the information exchange over the network.
The thick solid arrows represent the sequence of the called methods,
the thin solid arrows represent the required inputs,
and the dotted arrows represent the outputs (or inputs provided to subsequently called methods).
The pipeline part highlighted with the gray outline can be executed multiple times.}
\label{f:spux-contract-scheme}
\end{figure}

The schemes outlining designs of the SPUX executor (the manager side)
and the corresponding SPUX contract (the worker side)
are listed in \autoref{f:spux-executor-scheme} and \autoref{f:spux-contract-scheme}, respectively.
In \autoref{f:spux-executor-scheme},
during the configuration stage when executors are attached to SPUX components,
only the \code{setup} of each executor is called and only with the \code{owner} argument.
Such pre-setup executors are then used to compute the total required resources in framework \code{init(...)}.
After the configuration stage, the \code{init(...)} of the executor attached to the main component (\code{sampler})
is called by framework \code{init(...)} and initializes the connector (i.e. allocates computational resources).
The same executor \code{init(...)} method then starts a chain of recursive hierarchical initializations for all other executors.
In particular,
the \code{setup(...)} of the executor attached to the task of the owner of the current executor
is called, where the \code{task} of the executor owner
becomes the \code{owner} of the \code{task} executor,
and the \code{addresscall()} of the executor is used for
the \code{rootcall} of the \code{task} executor.
During the \code{bootup(...)}, each parallel worker is issued a contract
describing required worker behavior, as explained in \autoref{f:spux-contract-scheme}.
In \autoref{f:spux-contract-scheme},
the manager and peers communicators are provided as the arguments of the contract.
Initially, the port (to connect back to manager) and the executor (to be attached to incoming tasks)
for each worker's contract is received by the workers from the manager communicator.
The \code{taskroot} (integer or string) is determined by the \code{rootcall(...)} of the task executor,
which is set in \code{task.executor.setup (...)}, see \autoref{f:spux-executor-scheme}.
The \code{taskport} (integer or string) is determined by calling \code{init(...)} of the received task executor,
completing this way the next recursive step of hierarchical executor initializations.
The contract then waits for further instructions from the manager, for instance, to execute some tasks.
The determined \code{taskroot} and \code{taskport}
are unique among all other peer task executors (initialized in the contracts of parallel workers at the same hierarchical level),
and hence can be used for binding any requested task to the task executor initialized specifically for this contract.
The results obtained by executing the received tasks are then sent back to the manager.

\end{document}